\documentclass[letterpaper,twoside,10pt]{article}
\usepackage{amssymb}
\usepackage{amsmath}
\usepackage{latexsym}
\usepackage{verbatim}
\usepackage{graphicx}
\usepackage{epsfig}
\usepackage{stmaryrd}
\usepackage{rotating,graphicx}
\usepackage[english]{babel}
\usepackage{setspace}
\usepackage[english]{babel}
\usepackage[utf8]{inputenc}
\usepackage[T1]{fontenc}
\usepackage{lmodern}
\usepackage{pstricks}
\usepackage[round]{natbib}
\newcommand{\bra}{\langle}
\newcommand{\ket}{\rangle}
\newcommand{\kets}[1]{\left\vert #1 \right\rangle}% |1>
\newcommand{\bras}[1]{\left\langle #1 \right\vert}% <1|

\begin{document}

\begin{large}
\begin{center}
\textbf{{\Large The weight of collapse: dynamical reduction models in general relativistic contexts}}
\end{center}
\end{large}

\begin{center}
%Elias Okon and Daniel Sudarsky
Elias Okon %\\[.15cm]

\textit{Instituto de Investigaciones Filos\'{o}ficas, Universidad Nacional Aut\'{o}noma de M\'{e}xico\\
Circuito Maestro Mario de la Cueva s/n, Distrito Federal, 04510, Mexico}\\
\textit{E-mail:} \texttt{eokon@filosoficas.unam.mx} \\ [.45cm]

Daniel Sudarsky %\\[.15cm]

\textit{Instituto de Ciencias Nucleares, Universidad Nacional Aut\'{o}noma de M\'{e}xico\\
Apartado Postal 70-543, Distrito Federal, 04510, M\'{e}xico}\\
\textit{E-mail:} \texttt{sudarsky@nucleares.unam.mx} \\[.45cm]

\end{center}
\noindent \textbf{Abstract:} 
Inspired by possible connections between gravity and foundational question in quantum theory, we consider an approach for the adaptation of objective collapse models to a general relativistic context. We apply these ideas to a list of open problems in cosmology and quantum gravity, such as the emergence of seeds of cosmic structure, the black hole information issue, the problem of time in quantum gravity and, in a more speculative manner, to the nature of dark energy and the origin of the very special initial state of the universe. We conclude that objective collapse models offer a rather promising path to deal with all of these issues.

%\tableofcontents
%%%%%%%%%%%%%%%%%%%%%%%%%%%%%%%%%%%%%%%%%%%%%%%%%%%%%%%%%%%%%%
%%%%%%%%%%%%%%%%%%%%%%%%%%%%%%%%%%%%%%%%%%%%%%%%%%%%%%%%%%%%%% 
\section{Introduction}
\label{Int}
%%%%%%%%%%%%%%%%%%%%%%%%%%%%%%%%%%%%%%%%%%%%%%%%%%%%%%%%%%%%%%
%%%%%%%%%%%%%%%%%%%%%%%%%%%%%%%%%%%%%%%%%%%%%%%%%%%%%%%%%%%%%% 
The project of constructing a quantum theory of gravity is often regarded as entirely independent of the one devoted to clarifying foundational questions within standard quantum theory---the latter, a task mostly performed in the non-relativistic domain. There are, however, very suggestive indications that the two topics are intimately connected. To begin with, the standard interpretation of quantum mechanics, crucially dependent on the notions of \emph{measurement} or \emph{observer}, seems ill-suited to be applied to non-standard contexts, such as radiating black holes or the early universe. With this kind of scenarios in mind, observer-independent versions of quantum mechanics, such as Bohmian mechanics (\cite{Bohm}) or objective collapse models (\cite{GRW,CSL}), appear to be better options.\footnote{Everettian theories inspired by \cite{Eve:57} and the consistent histories approach developed in \cite{CH1,CH2,CH3} also aim at constructing observer independent formalisms. Unfortunately, at least for now, both of these programs seem to be plagued by insurmountable problems (see e.g., \cite[sec. 4]{MW} and \cite{CHus1,CHus2,CHus3}).} Moreover, throughout the years, people like R. Penrose or L. Diosi, among others, have uncovered intriguing connections between quantum-foundational issues and the quantum-gravity interface that are worth taking seriously (see, e.g., \cite{Mie:74,Penrose1,Diosi1,Ell:89,Per:95,Mass-dep1,Gol.Teu:01,Ryu:06,EREPR}). 

In this paper we describe a line of research\footnote{Papers where this project has been developed include: \cite{P1,P2,P3,P4,P5,P6,P7,P8,P9,P10,P11,P12,P13,P14,P15,P16}.} strongly influenced by this type of ideas and show that it has substantial promise in addressing various open issues confronting, not only the quantum-gravity interface in general, but also questions usually thought to belong solidly to the domain of quantum gravity research. The program explores the way in which objective collapse models (also know as dynamical reduction theories) can be used in dealing with a list of open issues in cosmology and quantum gravity. In particular, we have argued that objective collapse theories can help:
\begin{enumerate}
\item Solving a critical conceptual problem with the inflationary account of the emergence of seeds of cosmic structure.
\item Revising the expectation that inflation will give rise to, by now detectable, B-modes in the CMB.
\item Explaining dark energy in terms of energy non-conservation.
\item Diffusing the black hole information puzzle.
\item Dealing with the problem of time in quantum gravity.
\item Shedding light on the origin of the second law of thermodynamics and the very special initial state of the universe.
\end{enumerate}

An obvious initial problem with all this, though, is that most well-developed collapse models, such as GRW of CSL, are non-relativistic; and while at least three relativistic versions have been proposed recently (\cite{Tumulka,Bed,PearleRel}), none of them, as they stand, seems adequate for a general relativistic context. Therefore, as we will see below, part of the project we are describing has focused on developing methods and ideas in order to adapt objective collapse models for scenarios involving gravity.

Before getting into the details of this and related issues, it is convenient to say a few words regarding the philosophy behind the research line we have followed, in which we tackle the exploration of the interface between general relativity and quantum theory in a \emph{top-bottom} approach. In the usual \emph{bottom-up} quantum gravity programs (e.g., string theory, loop quantum gravity, causal sets, dynamical triangulations, etc.) one starts by assuming one has constructed a fundamental theory of quantum gravity and attempts to connect it to regimes of interest in the ``world out there,'' such as cosmology or black holes. In the \emph{top-bottom} approach, in contrast, one tries to push existing, well-tested theories to address open issues that seem to lie just beyond their domain of applicability---and in doing so, one considers introducing suitable modifications of the theoretical framework. The hope is for this latter approach to lead to the formulation of effective theories that could provide clues about the nature of more fundamental theories, such as quantum gravity. We see this path as emulating the development of quantum theory itself, which started with {\it ad hoc} additions to the classical theory that where latter employed in the construction of the general formalism. The idea, then, is to push general relativity and quantum field theory in curved spacetime into realms often deemed to be beyond their reach and to consider questions usually not explored in such contexts.
 
This manuscript is organized as follows. In section \ref{GandQM} we present a brief discussion of possible connections between gravity and foundational question in quantum theory. Section \ref{GRandC} is devoted to the task of adapting objective collapse models to a general relativistic setting. In section \ref{App} we consider the application of these ideas to inflationary cosmology, the dark energy issue, the black hole information puzzle, the problem of time in quantum gravity and the problem of explaining the special nature of the initial state of the universe. We wrap up with some final thoughts in section \ref{D}.
%%%%%%%%%%%%%%%%%%%%%%%%%%%%%%%%%%%%%%%%%%%%%%%%%%%%%%%%%%%%%% 
%%%%%%%%%%%%%%%%%%%%%%%%%%%%%%%%%%%%%%%%%%%%%%%%%%%%%%%%%%%%%% 
\section{Gravity and the foundations of quantum mechanics}
 \label{GandQM}
%%%%%%%%%%%%%%%%%%%%%%%%%%%%%%%%%%%%%%%%%%%%%%%%%%%%%%%%%%%%%%
%%%%%%%%%%%%%%%%%%%%%%%%%%%%%%%%%%%%%%%%%%%%%%%%%%%%%%%%%%%%%% 
As has been recounted many times before, one of the most notable moments during the Einstein-Bohr debates occurred when Bohr surprisingly used considerations based on general relativity in order to counter one of the best challenges Einstein managed to present against the, then new, quantum theory (see \cite{Bohr}). That certainly was a sensational knock out by Bohr, which not only showed that he was a serious intellectual contender for Einstein, but contributed to the widespread perception that, regarding the general discussion concerning the suitability of quantum mechanics as a basic theory of the micro-world, Bohr carried the day. The resurgence of interest in foundational questions of quantum theory, largely motivated by the work of J. Bell (see \cite{BellB}) and reflected by the increasing number of conferences and workshops devoted to the subject (and, indeed, the fact that this very book is being written), attest to the fact that Bohr's victory that day was not the end of the story.

As we mentioned in the introduction, even though (in spite of Bohr's move!) work in quantum gravity is often regarded as independent of foundational questions within quantum theory, there are very suggestive indications that the two issues might be intimately connected. Of course, chief among the conceptual problems of quantum theory is the so-called \emph{measurement problem}, which roughly corresponds to the fact that the standard formulation of quantum theory crucially depends on the concept of measurement, even though such notion is never formally defined within the theory. With standard laboratory applications in mind, this might not seem that problematic (one might consider, for instance, adopting an instrumentalist point of view). However, when the system under study is the whole universe, the inadequacies of such an approach become quite evident: in such case, there is nothing outside the system that could play the role of an observer. It is clear, then, that if one wants to apply quantum theory in such scenarios, it is essential to employ an alternative to standard quantum mechanics designed to deal with the measurement problem (see \cite{Bel:81,Hartle-Cosmology} for similar assessments).
 
Despite the fact that many interesting alternatives to the standard interpretation of quantum mechanics have been proposed, a fully satisfactory solution to the measurement problem, particularly one that is adequate for the relativistic domain, has not yet been found.\footnote{In fact, as argued in \cite{Sorkin}, one faces serious problems just by attempting to extend the standard interpretation of quantum theory to the quantum field theory context.} Such fact has not precluded quantum mechanics of displaying a spectacular predictive power. This, in turn, has led to a widespread pragmatic attitude in physics regrading the conceptual problems of quantum mechanics and, naturally, people working on different quantum gravity programs, such as string theory or loop quantum gravity, have inherited this attitude. Nevertheless, it is quite possible for the solutions to both of these issues to be related: since we already have managed to develop successful quantum theories of all other known interactions, the solution to the measurement problem may well lie in the quantum theory which is still lacking, that is, quantum gravity. On the other hand, the numerous failed attempts to construct a theory of quantum gravity, working under the assumption that the standard interpretation is correct, suggest that it may be necessary to solve the measurement problem in order to achieve the goal of building a quantum theory of gravity.

Intriguing connections between quantum-foundational issues and the quantum-gravity interface have been pointed out throughout the years. For instance, in \cite{Penrose1}, Penrose considers a large box containing a black hole in thermal equilibrium with a surrounding environment and argues that, since different initial states give rise to the same final one (due to the no-hair theorems), the black hole causes trajectories in phase space to converge. He claims, however, that this loss of phase space volume has to be compensated by some kind of bifurcation in the behavior of the non-black hole region and suggests that such bifurcations could be identified with the multiplicity of possible outcomes resulting from a quantum measurement (where several outputs may follow from the same input). This, together with alleged intrinsic spacetime instabilities when macroscopic bodies are placed in quantum superpositions of different locations, lead him to conjecture in \cite{Penrose2} a link between quantum collapses and gravity. 

Following a similar line of thought, in \cite{Diosi1,Diosi2} it is noted that solutions to the Newton-Schrödinger equation that represent quantum particles as self-gravitating objects could, in turn, be natural states that result from a kind of spontaneous self-localization due to the collapse of the wave function. More recently, in \cite{DiosiNew}, an interesting proposal has been put forward that seems to allow for a self-consistent, semi-classical treatment in which classical, Newtonian gravity is sourced by the collapsing quantum states of matter (see also \cite{Der:14,Nim.Hor:15}). Additionally, in \cite{Mass-dep1} it is suggested for the gravitational curvature scalar to be the field that causes the spontaneous collapses in objective collapse models and in \cite{Gol.Teu:01} it is argued that the major conceptual problems of canonical quantum gravity, namely, the problem of time and the problem of diffeomorphism invariant observables, are automatically solved by employing a Bohmian version of quantum mechanics.

Another recent work in which connections between gravity and quantum foundations are allegedly uncovered is \cite{Pik:15}, where it is argued that a universal, gravity-induced, type of decoherence explains the quantum-to-classical transition of systems with internal degrees of freedom. However, in \cite{Bon:15,Bon:16}, we rebut the claim that gravitation is responsible for the reported effect, we contest the fact that the effect is universal and, finally, we challenge the ability of decoherence to explain the quantum-to-classical transition.\footnote{See \cite[sec. 2]{P15} for a deeper analysis of why decoherence does not explain the quantum-to-classical transition nor helps in the resolution of important foundational problems in quantum theory.} From all this we conclude that gravity does not help account for the emergence of classicality as claimed.

An important issue within the foundations of quantum mechanics is the ontological question of `what is reality made of according to quantum mechanics?' (see for instance \cite{WF}). It is worth noting that answers to such question are often framed in a context where the existence of a classical spacetime is taken for granted. That is, things are postulated to populate a classical spacetime, which is assumed to exist independently of the rest of the stuff that the theory is supposed to refer to (see, e.g., \cite{Allori}). Of course, in contexts where one is considering a quantum description of spacetime itself, the issue can no longer be addressed in this standard form (for example, in such a context, one cannot only say that what exists are Bohmian particles or a mass distribution, as both of them require a background spacetime to give them their full meaning). The point is that discussions regarding ontology within a quantum gravity context will probably require a substantial reformulation and even a change of language and concepts before they can be fully addressed. It is clear, though, that at the level of the discussion of the present work, immersed in a semi-classical setting, we need to content ourselves with one of the various ontological pictures developed in a non-gravitational context and take it only as a partial and effective answer to the issue.

It is worth noting at this point that, in contrast to what is often assumed, the interface between quantum theory and gravitation need not involve the Planck regime. Consider, for instance, trying to describe the spacetime associated with a macroscopic body in a quantum superposition of being in two distinct locations. Clearly, such scenario implicates both gravity and quantum mechanics, without ever involving Planck scale physics. In fact, \cite{Page} describes an experiment devised to explore precisely such situation whose results are taken as evidence against semi-classical general relativity. First, they argue that if quantum states undergo standard (Copenhagen-style) collapses, then the semi-classical Einstein equations are inconsistent because during collapses the energy-momentum tensor is not conserved. That leads them to adopt an Everettian no-collapse interpretation which, however, they claim is in conflict with the performed experiment. Either way, they conclude that semi-classical general relativity is not viable. Another well-known argument against a semi-classical theory is offered in \cite{EH}, where a thought experiment is used to defend the claim that a semi-classical scenario leads to violations of either momentum conservation or the uncertainty principle, or to faster-that-light signaling. 

The above results, like many other in the literature that try to reach the same conclusion, namely, that semi-classical general relativity is not an option, seem rather strong. However, as has been shown repeatedly in, e.g., \cite{CH,Mat1,Mat2,Carlip}, none of them is really conclusive. For example, regarding \cite{Page}, it is clear that its argument relies on Copenhagen or Many Worlds being the only alternatives for interpreting the quantum formalism, but neither seems to be viable for the scenario considered. As for \cite{EH}, it has been shown that the proposed devise is impossible to build and, even if built, would not actually lead to a no-go theorem for a semi-classical formalism. In this project we work in a semi-classical setting; however, we do so keeping in mind that such description should not be taken as fundamental, but only as a good approximation, valid under appropriate circumstances (see, e.g., \cite{Kuo}).\footnote{In fact, most attempts at connecting bottom-up approaches to quantum gravity with ``real world'' scenarios rely on a semi-classical approach, in which matter is described quantum mechanically but spacetime is described classically.} As we see below, this allows us not only to address a number of open issues is cosmology and quantum gravity, but also to search for clues regarding a more fundamental description of spacetime.

Most people working on quantum gravity expect general relativity to be replaced by a theory of quantum gravity that will, among other things, ``cure the singularities'' of the former. However, our current understanding regarding how this replacement is going to work is still filled with serious gaps that offer enough room to accommodate a wide range of ideas. A key unresolved conceptual difficulty in quantum gravity research is the problem of recovering a classical spacetime out of a theory that does not contain such a concept at the fundamental level. Many people have suggested that spacetime could be an emergent phenomenon (\cite{JacobsonEmergent,SeibergEmergent,SorkinEmergent} and many others) and, if so, just as the thermodynamic notion of heat has no clear counterpart at the statistical mechanics level of description, spacetime concepts could only become meaningful at the classical level; and even in some of the schemes where the geometric degrees of freedom do appear at the fundamental level (e.g., causal sets and loop quantum gravity), the recovery of standard spacetime concepts often is possible only at the level of large aggregated systems.

The important point for us is that, under these scenarios, it would be unreasonable to expect the Einstein equations to provide a description of spacetime that goes beyond a phenomenologically useful characterization. Instead, we should think of the relativistic description of spacetime in analogy with, say, a hydrodynamic description of a fluid in terms of the Navier-Stokes equations. In such a case, it is clear that the offered description is no more than a good approximation, which does not contain the fundamental degrees of freedom and breaks down under certain circumstances (i.e., turbulence or the break of a wave). Therefore, we should not be unduly surprised if, in association with the quantum gravity interface, we encounter situations in which the Einstein equations fail to be satisfied exactly. In fact, there might be situations in which the metric characterization of spacetime provides a good description, which is, however, punctuated by small interruptions associated with the onset of phenomena that cannot be fully accounted for at that level of description. We submit that this kind of situation is precisely what one finds during quantum collapses, where the semi-classical description simply breaks down. However, instead of abandoning the project at that point, and in tune with our top-bottom approach, we propose to treat such breakdowns phenomenologically, within a trial an error frame of mind that seeks to extract clues about a suitable description for the more fundamental degrees of freedom involved. That is, we propose to push forward by adding suitable adjustments, from which we hope to eventually obtain hints for the construction of a more fundamental theory of quantum gravity.\footnote{A somehow comparable approach is adopted in \cite{GP}, where, in order to deal with the recovery of time in quantum gravity, some dynamical variable is designated as a ``physical clock,'' relative to which wave functions for other variables are constructed. By doing so, it is found that one recovers only an approximate Schrödinger equation, with corrections that violate unitarity. The problem is that the precise scheme adopted in this and related works relies heavily on the notion of decoherence and, as such, suffers from rather serious problems discussed on general grounds in \cite{P15}. That of course does not mean that taking a slightly more general view of the subject could not lead to something that could lie at the bottom of collapse theories.} In the next section we explain in detail how we propose to do this.
%%%%%%%%%%%%%%%%%%%%%%%%%%%%%%%%%%%%%%%%%%%%%%%%%%%%%%%%%%%%%% 
%%%%%%%%%%%%%%%%%%%%%%%%%%%%%%%%%%%%%%%%%%%%%%%%%%%%%%%%%%%%%% 
\section{Objective collapse in a general relativistic setting}
 \label{GRandC}
%%%%%%%%%%%%%%%%%%%%%%%%%%%%%%%%%%%%%%%%%%%%%%%%%%%%%%%%%%%%%% 
%%%%%%%%%%%%%%%%%%%%%%%%%%%%%%%%%%%%%%%%%%%%%%%%%%%%%%%%%%%%%% 
One of the most promising proposals in order to deal with the measurement problem is the dynamical reduction or objective collapse program. Early examples of theories within such program include GRW (\cite{GRW}) and CSL (\cite {CSL}). The main idea of the project is to add non-linear, stochastic terms to the dynamical equation of the standard theory in order to achieve a unified description of both the quantum behavior of microscopic systems and the emergence of classical behavior (e.g., absence of superpositions) at the macroscopic level. Of course, the price to pay is a departure from Schrödinger's evolution equation, accompanied by a breakdown of unitarity. In contrast, one of the advantages of these theories is that they leave much of the structure of quantum mechanics intact, so that things as the EPR correlations are clearly accounted for. Moreover, they are susceptible to experimental testing and, in fact, a robust program for that is currently under way (see \cite{Bassi}).
 
As we mentioned above, most objective collapse models, such as GRW or the original version of CSL, are non-relativistic. However, recently, fully relativistic versions such as \cite{Tumulka,Bed,PearleRel} have been developed. The problem is that none of them, in their present form, is adequate for a general relativistic context. The objective of this section is to describe one particular way in which this could be achieved. Before doing so, though, we will quickly review the GRW and CSL models (see \cite{GRW} and \cite{CSL}, respectively). 

Non-relativistic GRW postulates each elementary particle to suffer, with mean frequency $\lambda_0$, sudden and spontaneous localization processes around judiciously chosen random positions. Such localizations are then shown to mimic the Born rule and to ensure a quick elimination of superpositions of well-localized macroscopic states. In \cite{Tumulka}, a relativistic version of the GRW model for \emph{non-interacting} Dirac particles is introduced. Such model works in Minkowski spacetime as well as in (well-behaved) curved background spacetimes.

The non-relativistic continuous spontaneous localization model, or CSL, is defined by a modified Schrödinger equation, whose solutions are given by
 \begin{equation}\label{CSL1}
 { |\psi,t\rangle_w = \hat {\cal T}e^{-\int_{0}^{t}dt'\big[i\hat H+\frac{1}{4\lambda_{0}}[w(t')-2\lambda_{0}\hat A]^{2}\big]}|\psi,0\rangle,}
\end{equation}
 with $\hat {\cal T}$ the time-ordering operator and $w(t)$ a random white-noise function, chosen with probability
 \begin{equation}\label{CSL2}
{PDw(t)\equiv{}_w \langle\psi,t|\psi,t\rangle_w \prod_{t_{i}=0}^{t}\frac{dw(t_{i})}{\sqrt{ 2\pi\lambda_{0}/dt}}}.
\end{equation}
As $t \to \infty$, the CSL dynamics inevitably drives the state of the system into an eigenstate of the operator $\hat A$. Therefore, it unifies the standard unitary evolution with a ``measurement'' of such an observable. In a non-relativistic setting, for the collapse operator $\hat A$ one usually chooses a smeared position operator.

In a relativistic setting, the natural thing to do is to assign a quantum state to every Cauchy hypersurface $\Sigma$ and to postulate a Schwinger-Tomonaga-type equation to implement the evolution from one such hypersurface to another hypersurface $\Sigma'$. For example, the evolution equation introduced in \cite{Bed} is
\begin{equation} \label{Tomonaga}
 d_x \Psi (\phi; \Sigma') = \left\lbrace -i J(x) A(x) d\omega_x - (1/2) \lambda_{0}^2 N^2(x)d\omega_x 
 +\lambda_{0} N(x) d W_x
 \right\rbrace\Psi(\phi; \Sigma) ,
 \end{equation}
where $ d\omega_x $ is the infinitesimal spacetime volume separating $\Sigma$ and $\Sigma'$, $ J(x)$ is an operator constructed out of matter fields, $W_x$ is a Brownian motion field and $A(x)$ and $N(x)$ are operators that modify the state of an auxiliary quantum field (for more details we remit the reader to \cite{Bed}). As for the value of $\lambda_{0}$ in all of these models, it has to be small enough in order to recover normal quantum behavior at a microscopic level, but big enough in order to ensure a rapid localization of macroscopic objects. This can be achieved by setting, for example, $\lambda_{0} \sim 10^{-16} sec ^{-1}$.

At this point it is important to make a few comments regarding the \emph{physical interpretation} of objective collapse theories. As we explained above, in order to solve the measurement problem, such theories avoid relying on the standard interpretation of the quantum state in terms of the Born rule. However, by removing such a probabilistic interpretation, \emph{without} substituting it by something else, one does not yet arrive at a proper physical theory capable of making empirically verifiable predictions because one lacks a translation rule between the mathematical formalism and the physical world the theory is supposed to describe. Therefore, objective collapse models require an alternative interpretation of the quantum state.

One option in this regard is to interpret the wave function directly as a physical field. However, it seems that by doing so one is obligated to also treat configuration space as physical (see \cite{Alb:96}). Another alternative, proposed in \cite{Bel:87}, is to take the GRW collapses as the objects out of which physical stuff is made. Yet another option, first presented in \cite{Ghi:95}, is to interpret the theory as describing a physical mass-density field, constructed as the expectation value of the mass density operator on the state characterizing the system; the relativistic version of this interpretation, with the energy-momentym tensor in place of mass density, is discussed in \cite{RevMat}. Notice that if one adopts the mass density ontology described above, then the semi-classical approach becomes much more natural because, within such an interpretation, the expectation value directly represents the energy and momentum distribution predicted by the theory.

Going back to our proposal, as we explained above, we will regard semi-classical general relativity as an approximate description with a limited domain of applicability, which must, however, be pushed beyond what is usually expected. In particular, we will incorporate quantum collapses to such a picture. It is clear, thought, that during collapses the expectation value of the energy-momentum tensor is not conserved. Therefore, at such events, the semi-classical Einstein equations 
\begin{equation}
\label{EE}
G_{\mu\nu}=8\pi G\langle\xi\vert \hat{T}_{\mu\nu}\vert\xi\rangle
\end{equation}
are not valid. The proposal, as we mentioned, is to see the issue as analogous to a hydrodynamic description of a fluid, in which the Navier-Stokes equations will not hold in some situations (e.g., when a wave is breaking), but can be taken to hold before and after the fact. Analogously, we will take the semi-classical Einstein equations not to hold during a collapse, but to do so before and after.

It is clear, then, that the approach requires some kind of recipe to join the descriptions just before and after a collapse. To do so, we use the notion of a \textit{Semiclassical Self-consistent Configuration} (SSC) introduced in \cite{P6}, which is defined as follows: the set $\lbrace g_{ab}(x),\hat{\varphi}(x), \hat{\pi}(x), {\cal H}, \vert \xi \rangle \in {\cal H} \rbrace$ is a SSC if and only if $\hat{\varphi}(x)$, $\hat{\pi}(x)$ and $ {\cal H}$ correspond a to quantum field theory for the field $\varphi(x)$, constructed over a spacetime with metric $g_{ab}(x)$, and the state $\vert\xi\rangle$ in $ {\cal H}$ is such that
\begin{equation}\label{scc}
G_{ab}[g(x)]=8\pi G\langle\xi\vert \hat{T}_{ab}[g(x),\hat{\varphi}(x)]\vert\xi
\rangle ,
\end{equation}
where $\langle\xi| \hat{T}_{\mu\nu}[g(x),\hat{\varphi}(x)]|\xi\rangle$ stands for the expectation value in the state $\vert\xi\rangle$ of the renormalized energy-momentum tensor of the quantum matter field $\hat{\varphi}(x)$, constructed with the spacetime metric $ g_{ab}$. In a sense, the above construction is the relativistic version of the Newton-Schrödinger system (\cite{Diosi1}), and just like it, it involves a kind of circularity in the sense that the metric conditions the quantum field, and the state of the latter conditions the metric. 

As we explain below, the SSC formulation allows us to incorporate quantum collapses of the matter sector into the semi-classical picture we are considering. The idea is for different SSC's to describe the situation before and after the collapse and for them to be suitably joined at the collapse hypersurface. In more detail, a spontaneous collapse of the quantum state that represents matter is accompanied by a change in the expectation value of the energy-momentum tensor which, in turn, leads to a modification of the spacetime metric. Note however that this demands a change in the Hilbert space to which the state belongs, so the formalism forces us to consider a transition from one complete SSC to another one, rather than simple jumps of states in one Hilbert space. That is, collpases should be characterized as taking the system from one complete SSC construction into another.\footnote{As noted in \cite{Prague}, this is connected with issues appearing in other approaches, such as the stochastic gravitation program of \cite{Hu-Verdaguer}.}

The concrete proposal to implement an objective collapse model within the SSC scheme is the following (for simplicity we consider a model with discrete collapses, akin to GRW). We start with an initial SSC, which we call SSC1, and we give rules to randomly select both a spacelike hypersurface $\Sigma_{C}$ of the spacetime of SSC1, on which the collapse takes place, and a \emph{tentative} or \emph{target} post-collapse state $ \vert \chi^{t} \rangle \in {\cal H}_1$. Such state is used below to determine $\vert\xi ^{(2)}\rangle $, the actual post-collapse state. Then, a new SSC, which we call SSC2, is constructed for the collapsed state to live in and the two SSC's are glued to form a ``global spacetime.'' To do so, the SSC2 construction is required to posses an hypersurface isometric to $\Sigma_C$, which serves as the hypersurface where the two spacetimes are joined (such condition is analogous to the Israel matching conditions for infinitely thin time-like shells). Finally, in order to construct $ \vert\xi ^{(2)}\rangle \in {\cal H}_2$ out of $ \vert \chi^{t} \rangle \in {\cal H}_1$, we demand that, on $\Sigma_C$,
\begin{equation}\label{scc-joining}
\langle\chi^{t}\vert \hat{T}^{(1)}_{ab}[g(x),\hat{\varphi}(x)]\vert\chi^{t}
\rangle =\langle\xi^{(2)}\vert \hat{T}^{(2)}_{ab}[g(x),\hat{\varphi}(x)]\vert\xi ^{(2)}\rangle ,
\end{equation}
where $\hat{T}^{(1)} $ and $\hat{T}^{(2)} $ are the renormalized energy-momentum tensors of SSC1 and SSC2, which depend on the corresponding spacetime metrics and the corresponding field theory constructions (see \cite{P6} for details).

Think again about the hydrodynamics analogy in which, before and after a wave breaks, the situation is accurately described by Navier-Stokes equations, but the breakdown itself is not susceptible to a fluid description. If we now take the limit in which the duration of the break tends to zero, we have two regimes susceptible to a fluid description, joined at an {\it instantaneous break}. The spacelike hypersurface $\Sigma_{C}$ that joins SSC1 and SSC2 is analogous to that. The intention of all this is to construct an effective description to resolve urgent issues and to eventually explore these matching conditions in search of clues about the underlying theory.

One extra element we have considered, initially in the context of black holes and later in cosmology, is the possibility for the collapse rate $\lambda_{0}$ to depend on the local curvature of spacetime. For example, one could have something like
\begin{equation}
\label{Wd}
\lambda \left( W \right) = \lambda_0 \left[ 1+\left( \frac{W}{\mu} \right)^\gamma \right]
\end{equation}
with $W$ some scalar constructed out of the Weyl tensor, $\gamma \geq 1$ a constant and with $\mu$ providing an appropriate scale. As we will see below, in the context of black holes, this dependence brings about all of the information destruction required in order to avoid a paradox and, in cosmology, it helps explain the very special initial state of the universe. An independent motivation for a curvature-dependent collapse parameter comes from the fact that studies of experimental bounds on such parameter suggest it must depend on the mass of the particle in question (\cite{Mass-dep2,Mass-dep1}). Having the parameter to depend on curvature seems like a attractive way to implement this in a general relativistic context.

Finally, generalizing all this scheme to multiple collapses is straightforward. However, one must be careful of the fact that, unlike in a unitary quantum field theory, in which $\langle\xi_\Sigma \vert \hat{T}_{ab}(x) \vert\xi_\Sigma \rangle$ is independent of the hypersurface $\Sigma$, in collapse theories such an expression does depend on the choice of hypersurface. An interesting solution to this problem, spelled out in \cite{RevMat}, consists in stipulating that the hypersurface one must use in order to calculate $\langle\xi_\Sigma \vert \hat{T}_{ab}(x) \vert\xi_\Sigma \rangle$ has to be the past light cone of the point $x$. Finally, extending all this to the context of a continuous collapse theory, such as CSL, would involve some type of limiting procedure. While we do not expect such process to introduce extra conceptual difficulties, technically, it will surely be extremely demanding.
%%%%%%%%%%%%%%%%%%%%%%%%%%%%%%%%%%%%%%%%%%%%%%%%%%%%%%%%%%%%%% 
%%%%%%%%%%%%%%%%%%%%%%%%%%%%%%%%%%%%%%%%%%%%%%%%%%%%%%%%%%%%%% 
\section{Applications}
 \label{App}
%%%%%%%%%%%%%%%%%%%%%%%%%%%%%%%%%%%%%%%%%%%%%%%%%%%%%%%%%%%%%%
%%%%%%%%%%%%%%%%%%%%%%%%%%%%%%%%%%%%%%%%%%%%%%%%%%%%%%%%%%%%%%
Bellow we consider promising applications of the scheme described in the previous section to several open problems in cosmology and quantum gravity. 
%%%%%%%%%%%%%%%%%%%%%%%%%%%%%%%%%%%%%%%%%%%%%%%%%%%%%%%%%%%%%% 
\subsection{Cosmic inflation and the seeds of cosmic structure}
 \label{CI}
%%%%%%%%%%%%%%%%%%%%%%%%%%%%%%%%%%%%%%%%%%%%%%%%%%%%%%%%%%%%%% 
Contemporary cosmology includes \emph{inflation} as one of its most attractive components. Such process was conceived in order to address a number of naturalness issues in the standard big bang cosmological model, such as the horizon problem, the flatness problem and the exotic-relics problem (\cite{Liddle,Pea}). However, inflation's biggest success is to correctly predict the spectrum of the cosmic microwave background (CMB) and to account for the emergence of the seeds of cosmic structure.
 
The staring point of such an analysis is a flat Friedmann–Robertson–Walker (FRW) background spacetime, which is inflating under the influence of an homogeneous inflaton background field. On top of that, one considers perturbations of the metric and the inflaton, which are treated quantum mechanically and are initially set at an appropriate homogeneous vacuum state (essentially the so-called Bunch-Davies vacuum). Then, from the \emph{quantum fluctuations} of these perturbations, the primordial inhomogeneities and anisotropies are argued to emerge. These primordial inhomogeneities and anisotropies, in turn, are said to constitute the seeds of all the structure in our universe as, according to our current cosmological picture, they later evolved, due to gravitational attraction, into galaxy clusters, galaxies, stars and planets. The result of all this is a remarkable agreement between predictions of the inflationary model and, both, observations of the CMB and structure surveys at late cosmological times.
 
It is indeed remarkable that, with one exception, the theoretical predictions of inflation are in exquisite agreement with observations. The exception has to do with the fact that inflationary models, along with the formation of inhomogeneities and anisotropies, generically also predict the formation of primordial gravity waves, which should become manifest in the so-called B-modes in the CMB polarization (\cite{nullB-modes1,nullB-modes2,Kamionkowski}). The problem is that, up to this point, these B-modes have not been observed. We have more to say about this at the end of our analysis, but, in the meanwhile, we return to the central topic of the emergence of the seeds of structure. 

According to the standard picture, during inflation the universe was homogeneous and isotropic (both at the classical and quantum levels of description); in spite of this, the final situation was not homogeneous and isotropic: it contained the primordial inhomogeneities which resulted in the structure that, among other things, permitted our own existence. A natural question arises: how did this transition from a symmetric into a non-symmetric scenario happened, given that the dynamics of the whole system does not break those symmetries? The problem is that the quantum fluctuations from which the seeds of structure are supposed to emerge are not really \emph{physical} fluctuations but only a characterization of the \emph{width} of the quantum state. Therefore, those fluctuations are incapable of breaking the symmetries of the initial state. The standard account, then, implicitly assumes a transition from a symmetric into a non-symmetric state, without an understanding of the process that leads, in the absence of observers or measurements, from one to the other. Such fact renders the standard account unsatisfactory.
 
A similar issue was considered by N. F. Mott in 1929 concerning the $\alpha$ nuclear decay. In such scenario one starts with a $J=0$ nucleus undergoing a rotationally invariant interaction that leads to a state characterized by a spherical outgoing wave function. The problem is that what one observes in a bubble chamber are the straight paths of the outgoing $\alpha$ particles, which clearly break the spherical symmetry. It is often assumed that the issue was satisfactorily addressed at the time, but a close examination of that work shows this not to be the case. The problem is that the proposed solution was based on the assumption that atoms in the bubble chamber where, on the one hand, highly localized, thus breaking the symmetry of the complete system, and, on the other hand, taken to act as classical detectors with well-defined excitation levels (in contrast with suitable quantum superpositions thereof). Both of these issues are clearly related to the measurement problem we mentioned above. However, in the cosmological scenario, the problem presents itself in an aggravated form as, even if we accepted to give observers a fundamental role in the theory (something we do not do), one could not call upon such observers to play the role of inducers of quantum collapse in the very early universe, where no such beings existed.
 
It is worth noting that we face here a rather unique situation, where quantum theory, general relativity and observations come together. We should contrast the present case with other scenarios one might initially think explore similar realms, such as the COW experiment (\cite{COW}) or experiments that study neutron quantum states in earth's gravitational field (\cite{neutrons}). In fact, these experiments are only tests of the \emph{equivalence principle}, in the sense that the situations considered might be described in terms of falling reference frames and, when doing so, gravity simply disappears from the scene. In the inflationary context, the situation is radically different because the quantum state of the inflaton field and, in particular, the spatially dependent fluctuations, actually gravitate and affect things like the spacetime curvature. Therefore, in order to address the question at hand, namely the emergence of the seeds of cosmic structure, we need to call upon a physical process that occurs in time. After all, emergence actually \emph{means} for something not to be there at a time, but to be there at a latter time. In the present case, what we need is a \emph{temporal} explanation of the breakdown of the symmetry of the initial state; and the point we want to stress is that spontaneous objective collapse theories can accomplish this. What we propose, then, is to add to the standard inflationary paradigm an objective, spontaneous quantum collapse of the wave function in a form suitably adapted to the situation at hand (see \cite{P1,P2,P3,P4,P5,P6,P7,P8,P9,P16}). Below we develop these ideas in some detail.

%%%%%%%%%%%%%%%%%%%%%%%%%%%%%%% 
\subsubsection{Objective collapse and the seeds of cosmic structure} 
%%%%%%%%%%%%%%%%%%%%%%%%%%%%%%%
In the following we show how the implementation of a CSL dynamics resolves the issues we pointed out above regarding the standard account for the emergence of the seeds of structure during the inflationary era. In principle, all this should be done using the SSC formalism we described in section \ref{GRandC}. However, due to its complexity, we use instead a approximated scheme in which we always use the Hilbert space construction corresponding to the unperturbed spacetime metric (see below) and the collapse thus simply induces a jump into a different vector in the same Hilbert space. In \cite{P6} we have checked that, at the lowest order in perturbation theory, this is equivalent to the full-fledged SSC treatment. 

As in the standard account, we split the metric and the inflaton into a homogeneous background and a potentially inhomogeneous fluctuation, i.e., $g=g_0+\delta g, \phi=\phi_0+\delta\phi$. The metric background corresponds to a flat FRW universe, with line element $ ds^2=a(\eta)^2\lbrace- d\eta^2 + \delta_{ij} dx^idx^j\rbrace$ (with $\eta$ a conformal time), and the field background to a homogeneous scalar field $\phi_0(\eta)$. Such $\vec k = 0$ mode of the field, which is responsible for the overall inflationary expansion, is treated classically in this effective approximation. However, as shown in \cite{P6}, it can be treated quantum mechanically within the full SSC formalism. We here then concentrate on the $\vec k \not= 0$ modes of the field. Moreover, since our approach calls for quantizing the scalar field, but not the metric perturbation, we do not use the so-called Muckhanov-Sassaki variable as is customary in this field. 

The evolution equations for the background field and metric are given by
\begin{equation}
\ddot\phi_0 + 2 \frac{\dot a}{a}\dot\phi_0 + a^2V'(\phi_0) =0 \quad \text{and} \quad 3\frac{\dot a^2}{a^2}=4\pi G \left(\dot \phi^2_0+ 2 a^2 V(\phi_0)\right),
\end{equation}
with $V(\phi)$ the inflaton potential, `` $\dot{}$ '' $\equiv \partial/\partial\eta$ and `` $'$ '' $\equiv \partial/\partial\phi$. The solution for the scale factor corresponding to the inflationary era is $a(\eta)=-\frac{1}{H_{\rm I} \eta}$, with $H_I ^2\approx (8\pi/3) G V$ and with the scalar field $\phi_0$ in the slow roll regime, i.e., $\dot\phi_0= - ( a^3/3 \dot a)V'(\phi_0)$. As in the standard inflationary scenario, inflation is followed by a reheating period in which the universe is repopulated with ordinary matter fields. Such stage then evolves towards a standard hot big bang cosmology regime leading up to the present cosmological time. The functional form of $a$ during these latter periods changes, but we can ignore those details because most of the change in the value of $a$ occurs during the inflationary regime. We set $a=1$ at the present cosmological time and assume that the inflationary regime ends at a value of $\eta=\eta_0$, which is negative and very small in absolute terms.
 
Next one considers the perturbations. For the metric we write
\begin{equation}
ds^{2}=a^{2}(\eta)\lbrace -(1+2\Psi)d\eta^{2}+ [(1-2\Phi)\delta_{ij} + h_{ij }]dx^{i}dx^{j}\rbrace,
\end{equation}
where we are using the so-called longitudinal gauge. The scalar field, on the other hand, is treated using quantum field theory on such background, with its state $\vert\xi\rangle$ and the metric satisfying the semi-classical Einstein equations (\ref{EE}).
 
As we already explained, at the early stages of inflation (which we characterize by $\eta=-{\cal T}$) the state of the scalar field perturbation is described by the Bunch-Davies vacuum. As a result, spacetime at the time is totally homogeneous and isotropic and the operator $\delta\phi$ and its conjugate momentum $\pi_{\delta\phi}$ are characterized by Gaussian wave functions, centered on $0$, with uncertainties $\Delta { \delta\phi}$ and $\Delta{{\pi}_{\delta\phi}}$ (for ease of notation we omit the ``hats'' over the operators $\delta\phi$ and $\pi_{\delta\phi}$). Next enters the collapse, which randomly and spontaneously modifies the quantum state and, generically, changes the expectation values of $\delta\phi$ and $\pi_{\delta\phi}$. We assume that the collapse is controlled by a stochastic function, mode by mode. Such an educated guess will later be contrasted with observations. It is important to keep in mind that, in this picture, our universe corresponds to one specific realization of these stochastic functions (one for each $\vec k$). Thus, if we were given that specific realization, the prediction of what we would see in the CMB would be completely transparent and unambiguous. Given, however, that the theory is fundamentally stochastic, we have no way of {\it a priori} determining the specific realization of such functions and thus, in order to make concrete predictions, we need to resort to some statistical manipulations. We come to this issue shortly. 
 
From equation (\ref{EE}) and what we have said so far we obtain
\begin{equation}
\nabla^2\Psi= 4 \pi G \dot{\phi_0} \bra \dot{\delta\phi}\ket ,
\end{equation}
from which it follows that, as soon as the expectation value of $\dot{\delta\phi}$ deviates from zero, which generically will due to the spontaneous collapses, so will the metric perturbation $\Psi$. In the Fourier representation, the above equation reads
\begin{equation}\label{poisson}
-k^2 \Psi_{\vec k} = 4 \pi G \dot{\phi_0} \bra \dot{\delta\phi}_{\vec k}\ket = \frac{4 \pi G \dot{\phi_0}}{a}\bra \pi_{\vec k} \ket .
\end{equation}

In order to derive observational consequences of all this, we note that there is a direct connection between $\Psi_{\vec k}$ and the temperature fluctuations of the CMB observed today. In fact we have
 \begin{equation}\label{cmbB1}
 \frac{\Delta T(\theta, \varphi)}{\bar T}= c \int d^3 k e^{i\vec{k}\cdot \vec{x}}
 \frac{1}{k^{2}}\bra \hat \pi_{\vec k}(\eta_D)\ket \quad \text{with} \quad c\equiv -\frac{4 \pi G \dot{\phi}_0(\eta)}{3a},
\end{equation} 
where $\vec{x}$ is a point on the intersection of our past light cone with the last scattering surface ($\eta= \eta_D$) corresponding to the direction on the sky specified by $\theta, \varphi$. Decomposing in spherical harmonics, we obtain
\begin{equation}\label{cmbB2}
\alpha_{lm}= c \int d^{2}\Omega Y_{lm}^{*}(\theta, \varphi) \int d^3 k e^{i\vec{k}\cdot \vec{x}} \frac{1}{ k^2}\bra \hat \pi_{\vec k}(\eta_D)\ket .
\end{equation} 
It is worthwhile pointing out that, within the standard approach, the expression analogous to the last equation is never shown. That is because the prediction in such approach for this quantity is in fact zero. Of course, what practicing cosmologists do at this point is to bring some loosely worded arguments explaining that quantum theory should not be used to predict what we see directly but only for predicting averages. Therefore, they argue, one should instead consider the quantity
 \begin{equation}\label{Cl}
 C_l \equiv \frac{1}{2l+1} \Sigma_{m}|\alpha_{lm}|^2 ,
\end{equation} 
which characterizes the orientation average of the magnitude of the $a_{lm}$. This line of thought raises several questions: why does the theory only make predictions for the orientation average and not for the $a_{lm}$ directly?, or why should we ignore the fact that we actually observe the temperature at each pixel in the sky and not the average? We do not think the standard approach offers good answers for these questions. Within the approach we are considering, in contrast, the conceptual picture is clear, with a straightforward explanation for where the randomness occurs, what is its source, and how it plays a role in yielding predictions. 

In more detail, within our approach, the quantity in equation (\ref{cmbB2}) can be thought of as a result of a \emph{random walk} on the complex plane, each step of the walk determined by the random function controlling the CSL evolution of mode $\vec k$, and the integration representing the addition of the infinitesimal contributions of each mode to the complete walk. One of course cannot predict the end point of such walk, but one can instead focus on the magnitude of the total displacement to compute $|\alpha_{lm}|^2$ and estimate such value via an ensemble average 
\begin{equation}
\label{ave}
\overline {\bra \pi_{\vec{k}}(\eta)\ket \bra \pi_{\vec{k}'}(\eta)\ket^*} = f(k) \delta( \vec{k}-\vec{k}').
\end{equation}
It can be checked that agreement between predictions and observations requires that $ f(k) \sim k $.

In order to compute the average in (\ref{ave}) using CSL we work with a rescaled field $y (\eta, \vec x) \equiv a \delta\phi (\eta, \vec x) $ and its momentum conjugate $\pi_y (\eta, \vec x) = a \delta\phi'(\eta, \vec x) $. We also put everything in a box of size $L$ (to be removed at the end) and focus on a single mode $ \vec k$, so we write
\begin{equation}
{ \hat{Y} \equiv (2\pi/L)^{3/2} y (\eta, \vec k), \qquad 
 \hat{\Pi} \equiv (2\pi/L)^{3/2}
 \pi_y (\eta, \vec k)}.
\end{equation} 
What we need to do next is to evaluate the ensemble average $\overline{\langle \hat{\Pi} \rangle^{2}}$ and determine under what circumstances, if any, it behaves as $\sim k$. Before doing so we note that, as shown in the expression above, we must consider $\overline{\langle \hat{\Pi} \rangle^{2}}$ and not the quantity $\overline{\langle \hat{\Pi}^2\rangle}$, which is the focus of attention in traditional approaches. One of the reasons for doing so is made evident by the observation that the latter is generically non-vanishing even for states that are homogeneous and isotropic, such as the vacuum state, while the former will vanish when the state under consideration is the vacuum---thus clearly exhibiting the fact that no anisotropies can be expected in fully isotropic situations. 

In order to apply CSL to compute the required average, we need to choose the operator $\hat A$ driving the collapse. In this regard we first assume that the operator driving the collapse of the mode $\vec k$ is the corresponding operator $\hat{\Pi}$. That is, we set $\hat A=\hat{\Pi}$ in the CSL evolution equation (\ref{CSL1}). The calculation is straightforward although quite long so we refer the reader to \cite{P7} for details. In the end one obtains
\begin{equation}\label {P27}
\overline{\langle \hat{\Pi}\rangle^{2}}= \frac{\lambda k^{2}{\cal T}}{2} + \frac{k}{2} - \frac{k}{\sqrt{2}\sqrt{ 1 +\sqrt{1+ 4\lambda^2}}}.
\end{equation}
\noindent 
As we mentioned above, in order to obtain results consistent with CMB observations, we need $\overline{\langle \hat{\Pi}\rangle^{2}}$ to be proportional to $k$. We can achieve this if we assume that the first term is dominant and that
\begin{equation}
\lambda=\tilde\lambda/k
\end{equation}
with $\tilde\lambda$ a constant independent of $k$. Note that this replaces the dimensionless collapse rate parameter $\lambda$ with $\tilde\lambda$ having dimensions of time${}^{-1}$. Doing this we obtain
\begin{equation}
{ \overline{\langle \hat{\Pi}\rangle^{2}}=\frac{\tilde \lambda k{\cal T}}{2} + \frac{k}{2} - \frac{k}{\sqrt{2}\sqrt{ 1 +\sqrt{1+ 4(\tilde \lambda/k)^2}}}}.
\end{equation}

Analogously, if instead of $\hat \Pi$ as the generator of collapse, we take $\hat Y$, we obtain
\begin{equation}
 \overline{\langle \hat \Pi \rangle^{2}} 
 =\frac{\lambda {\cal T}}{2} + \frac{k}{2}
 \left\lbrace 1- \frac{(1 + 4(\lambda/k^2)^{2})}{ F( \lambda/k^2)+ 2 ( \lambda/k^2)^2 F^{-1}(\lambda/k^2) - 2( \lambda/k^2)(k \eta)^{-1} } \right\rbrace ,
\end{equation}
where $F(x) \equiv \frac{1}{\sqrt{2}}\sqrt{1 + \sqrt{1 + 4x^2 }}$.% ($F(0)=1$). 
%Once more, if we turn off CSL, $\lambda=0$ we find $\overline{\langle \hat \Pi \rangle^{2}}=0$.
Therefore, in order to obtain results consistent with observations, we need to assume that the first term dominates and that
\begin{equation}
\lambda=\tilde\lambda k.
\end{equation}
This time the collapse rate parameter $\lambda$ of dimension { time$^{-2}$ is replaced with the parameter $\tilde\lambda$ of dimension time$^{-1}$. Doing this leads to
\begin{equation}
{ \overline{\langle \hat \Pi \rangle^{2}}=\frac{\tilde \lambda k{\cal T}}{2} + 
 \frac{k}{2} \left\lbrace1-
 \frac{(1 + 4(\tilde {\lambda}/k)^{2})}{ F( \tilde {\lambda}/k)+ 2 ( \tilde {\lambda}/k)^2 F^{-1}(\tilde {\lambda}/k) - 2( \tilde {\lambda}/k)(k \eta)^{-1} }\right\rbrace }. 
\end{equation}
% The resulting amplitude for the scalar perturbation spectrum is then :...
 
Finally, comparisons with observations, using the GUT scale for the value of the inflation potential and standard values for the slow-roll parameter, leads to an estimate of $\tilde\lambda \sim 10^{-5} MpC^{-1} \approx 10^{-19} sec ^{-1}$. The fact that this is not very different from the GRW suggested value for the collapse rate is indeed encouraging.
%%%%%%%%%%%%%%%%%%%%%%%%%%%%%%% 
\subsubsection{Tensor modes}
\label{Tm}
%%%%%%%%%%%%%%%%%%%%%%%%%%%%%%% 
As we mentioned above, the theoretical predictions of inflation are in exquisite agreement with observations, with one exception. Inflationary models generically predict the production of B-modes in the CMB, but these B-modes have not been observed. Such fact has been used to severely constrain the set of viable inflationary models (see \cite{nullB-modes1,nullB-modes2,Kamionkowski}). Here we show that the incorporation of objective collapses into the picture dramatically alters the prediction for the shape and size of the B-mode spectrum, explaining why we have not seen them.

Within the standard approach, both the scalar and tensorial perturbations are treated equally. It is not surprising, then, that standard inflationary models predict a precise relationship between their amplitudes and shapes. In fact, the standard estimates for the power spectra of the scalar and tensor perturbations are given by
\begin{equation}
P_S^2 (k) \sim \frac {1}{k^3}\frac{V}{M_p^4 \epsilon} \quad \text{and} \quad
P_h^2 (k) \sim \frac {1}{k^3}\frac{V}{M_p^4 } 
\label{usual-exp}
\end{equation}
respectively and direct measurements of $\epsilon$ have been used to limit the viability of some of the simplest inflationary models. Above we showed that adding collapses into the picture solves a grave conceptual problem for the standard account, without modifying the prediction for the scalar power spectrum. Bellow we show that the story is different with the tensor perturbations.

From (\ref{EE}), the equation of motion for the tensor perturbations is
\begin{equation}
(\partial^2_0-\nabla^2)h_{ij}+2 (\dot a/a)\dot h_{ij} =16\pi G \langle (\partial_{i} \delta \phi)(\partial _{j}\delta \phi) \rangle^{tr-tr} ,
\label{gw}
\end{equation}
where $tr-tr$ stands for the transverse traceless part of the expression. In terms of a Fourier decomposition, we need to solve the equation,
 \begin{equation}
\ddot { h}_{ij} (\vec k, \eta)+2 (\dot a/a)\dot { h}_{ij} (\vec k, \eta) + k^2 { h}_{ij} (\vec k, \eta) =S_{ij} (\vec k, \eta) ,
\label{gw1}
\end{equation}
with zero initial data, and source term given by
\begin{equation}
S_{ij} (\vec k, \eta) =16\pi G \int \frac{d^3x}{\sqrt{(2\pi)^{3}}} e^{i\vec k \vec x} \langle (\partial_{i} \delta\phi)(\partial_{j}\delta\phi) \rangle^{tr-tr} (\eta, \vec x).
\end{equation}
Formally, this is a divergent expression, but we must introduce a cut-off $P_{UV}$ that one might take as the scale of diffusion dumping.% (\cite{damp}). 
Doing so leads to our prediction for the power spectrum of tensor perturbations
\begin{equation}
\mathcal{P}_h^2 (k) \sim \frac {1}{k^3}\left(\frac{V}{M_p^4}\right)^2 \frac {P_{UV}}{k} ,
%\label{gw-Power}
\end{equation}
which is substantially smaller than the standard prediction above. Therefore, within this approach, we expected not to see tensor modes at the level they are being looked for. In fact, in order to have any possibility of detecting them, we would need to improve sensitivity by various orders of magnitude and to look for them at very large scales.
%%%%%%%%%%%%%%%%%%%%%%%%%%%%%%%%%%%%%%%%%%%%%%%%%%%%%%%%%%%%%% 
\subsection{Dark energy from energy and momentum non-conservation}
 \label{DE}
%%%%%%%%%%%%%%%%%%%%%%%%%%%%%%%%%%%%%%%%%%%%%%%%%%%%%%%%%%%%%% 
We have seen that a key issue to be confronted in incorporating collapse theories into the context of general relativity at the semi-classical level is the generic violation of the condition $ \nabla^a \langle T_{ab} \rangle =0 $. In order to deal with such a problem, we have taken semi-classical relativity as an effective description (analogous to that provided by the Navier-Stokes equations for fluid dynamics) and argued that collapses might be incorporated using the SSC formalism to join descriptions of before and after a collapse event. There is, however, a class of scenarios that can be dealt with using a slightly modified theory of gravity known as \emph{unimodular gravity}, a theory based on a traceless version of Einstein's equations
 \begin{equation}
R_{ab} - \frac{1}{4} g_{ab} R = 8\pi G \left( T_{ab} - \frac{1}{4} g_{ab} T \right).
\end{equation}

Unimodular gravity arises naturally when one considers reducing the general diffeomorphism invariance of general relativity to invariance under volume-preserving diffeomorphisms only. That, in turn, seems as a natural modification if one contemplates incorporating a ``constant rate of collapse events per unit of spacetime volume,'' as would be natural in a relativistic collapse theory. Moreover, in turns out that, in the unimodular version of general relativity, the energy-momentum conservation does not follow from the dynamical equations, so it has to be imposed by hand as an independent assumption. The point we want to make is that, with collapses in mind, we need not do that! It is important to remember, thought, that in this case the Bianchi identity leads to $ \nabla_{[c} \nabla^a T_{b]a} = 0 $, which is now the self-consistency condition for integrability of the equations.

Suppose, then, that we have a situation where we want to consider a collapse theory in a semi-classical setting, in which energy-momentum is (of course) not conserved, $ \nabla^a \langle T_{ab} \rangle \not= 0 $, but where $ \nabla_{[c} \nabla^a \langle T_{b]a} \rangle = 0 $. One such situation is in fact provided by cosmology, where the homogeneity and isotropy of the system guarantees the integrability condition is automatically satisfied. For the case of simply-connected spacetimes, such condition reduces to $ J_a =\nabla_a Q $ for some scalar quantity $Q$ (where we defined $ J_a \equiv \nabla^a \langle T_{ab} \rangle$) and, as shown in \cite{Josset:2016}, in that case one can recast the equations as
\begin{equation}
	R_{ab} - \frac{1}{2} R g_{ab} +\Lambda_\text{eff}
	 g_{ab}= \frac{8 \pi G}{c^4} \langle T_{ab} \rangle.
\end{equation}
where the effective cosmological constant is given by $\Lambda_\text{eff}= \left(\Lambda_0 +\frac{8 \pi G}{c^4} \int J \right)$, with $\Lambda_0 $ an integration constant. 
 
When energy-momentum is conserved, one of course recovers standard general relativity, with a cosmological constant given just as an integration constant. However, the interesting observation is that if one has some argument which allows one to fix the initial condition $\Lambda_0 $ at any time, together with an explicit mechanism for violation of energy and momentum, one can estimate the value of the effective cosmological constant at other times. Of particular interest in this regard is the unexpected value for the cosmological constant needed to explain the late accelerated expansion of the universe discovered close to two decades ago (\cite{Riess:1998cb, Perlmutter:1998np}). The observed value of $\Lambda^\text{obs} \approx 1.1~10^{-52}~\text{m}^{-2}$ is unexpected because the seemingly natural values for $\Lambda$ are either zero or a value which is 120 orders of magnitude larger than the one favored by observations (see \cite{Weinberg:1988cp}).
 
In \cite{Josset:2016} it was noted that collapse theories can offer an attractive possibility in order to solve this puzzle. As is well-known, non-relativistic CSL with collapse parameter proportional to mass and with collapse operators given by smeared mass-density operators, leads to a spontaneous creation of energy that is proportional to the mass of the collapsing matter quantum state \cite{Mass-dep1, Mass-dep2}. In the cosmological setting, such an effect generates an energy-momentum violation given by
\begin{equation}
\label{EnergyIncreaseCSL}
J = -\lambda_{ \text{CSL}} \rho^\text{b} dt,
\end{equation}
with $\lambda_{\text{CSL}}$ the CSL parameter and $\rho^\text{b}$ the energy-density of the baryonic contribution to the universe mean density (the contribution of lighter particles can be expected to be sub-dominant). Now, taking as initial time the era of hadronization (corresponding to $z_{\rm h}\approx 7 \,10^{11}$), one finds
% \eqref{Eff-Lambda} and \eqref{EnergyIncreaseCSL} lead to
\begin{equation}\label{DarkEnergyEstimate1}
\Lambda_\text{eff} -\Lambda_0 \approx - \frac{3 \Omega^\text{b}_0 H_0 \lambda_{ \text{CSL}}}{\sqrt {\Omega^\text{r}_0}c^2} z_{\rm h} \approx - \frac{\lambda_{ \text{CSL}}}{4.3\,10^{-31}\, \text{s}^{-1}} \Lambda^\text{obs},
\end{equation}
where standard values for the cosmological parameters where used (\cite{Adam:2015rua}). Given that the current allowed range for the CSL parameter is $3.3\,10^{-42} \text{s}^{-1} < \lambda_{\text{CSL}} < 2.8\,10^{-29} \text{s}^{-1}$, we arrive at a prediction which, remarkably, yields the correct order of magnitude.

It is very important to notice, thought, that this cannot be the whole story because the quantity estimated above has the opposite sign as the observed cosmological constant. On the other hand, such an estimate only contains contributions from baryonic matter in the late cosmological era, where the dominant form of matter is non-relativistic. We should also consider contributions from previous epochs, where the dominant forms of matter were relativistic. Moreover, in order to compare the theoretical estimate with observations, we need a condition fixing the value of $\Lambda_0$ at a suitable time. This latter input might possibly emerge from the inflationary scenario, where the spatially flat condition is an attractor of the dynamics, setting $ \Omega_{total } =1$, or from a suitable condition at the Planck scale. At any rate, the lesson we take from all this is that this perspective offers new paths for a possible reconciliation between reasonable theoretical predictions for the cosmological constant and its observed value.
%%%%%%%%%%%%%%%%%%%%%%%%%%%%%%%%%%%%%%%%%%%%%%%%%%%%%%%%%%%%%% 
\subsection{The black hole information issue}
 \label{BH}
%%%%%%%%%%%%%%%%%%%%%%%%%%%%%%%%%%%%%%%%%%%%%%%%%%%%%%%%%%%%%% 
According to general relativity, the end point of the evolution of any sufficiently massive object is a black hole. Furthermore, all black holes are expected to eventually settled into one of the small number of stationary back hole solutions. In fact, within the setting of the Einstein-Maxwell theory, these are fully characterized by just three parameters, namely, mass $ M$, charge $Q$ and angular momentum $J$. Work in theories including additional kinds of fields, such as scalar fields, have also led to so-called no-hair theorems that severely limit the possibility of enlarging this class of stationary black holes. Consideration of more general theories, such as those involving non-abelian Yang Mills fields, do lead to a new class of solutions. However, as it turns out, these new solutions represent unstable black holes. %\cite{BHYM}. 
Therefore, if we limit ourselves to the stable class of black holes as possible end points of the evolution, we face a situation where these final states are fully characterized by the three parameters $M$, $Q$ and $J$.

All this brings up the following issue. In general, an initial configuration that leads to the formation of a black hole, requires an extremely large amount of data to be fully characterized. However, as we just saw, the final state is fully described by only three parameters. What happens, then, to the information-preserving (i.e., deterministic) character of the laws of physics, which is supposed to allow us to use both initial data to predict the future and final data to retrodict the past? Well, if the black hole in question is eternal, as general relativity holds, then the above mentioned problem can be avoided by arguing that $M$, $Q$ and $J$ serve only to characterize the black hole's exterior, with all the remaining information about the initial state encoded in the interior. However, incorporating quantum aspects into the picture radically changes the situation.

As shown in \cite{Haw:75}, quantum field theory effects cause black holes to radiate. It is true that Hawking's calculation does not include back-reaction, but strong confidence in energy conservation bring people to the conclusion that, as Hawking's radiation takes away energy, the mass of the black hole has to diminish. Such loss of mass further increases the amount of radiation, apparently leading to the complete evaporation of the black hole in a finite time.\footnote{Quantum gravity effects could eventually stop the evaporation, leading to a stable remnant. However, given the low mass such objects would have, they are not expected to help in the resolution of the information loss issues (see below).} It seems that all this leads to a problem since, after the black hole evaporates completely, there is no longer an interior region to contain the initial information. Note however that the system contains a singularity ``into which information can fall,'' so still no proper inconsistency arises because it can be argued that the initial information merely escapes through the singularity. Remember, though, that quantum gravity is widely expected to cure such singularities, and if that is the case, then a conflict finally seems inevitable.

One possible way out of all this is to hold fast to the assumption that information is always conserved, in which case there are two options. First, one can assume that the Hawking radiation carries the information. However, as has been argued in \cite{firewalls}, that leads to the formation of a firewall on the horizon. Second, one can assume that the information is preserved in a remnant or in low energy modes that survive the evaporation. The problem with this second option is that one expects remnants or surviving modes to have an energy of the order of the Plank mass (because one expects Hawking's calculation to hold until the deep quantum gravity regime is reached), so it's really hard to envisage how they could encode an arbitrarily large amount of information.

The other alternative is to assume that information is indeed lost during the evaporation process, from which one could conclude that quantum mechanics needs modifications when black holes are involved. However, it is much more natural (and exciting!) to think of a more fundamental modification of quantum theory involving gravity. In such scenario, violations of unitarity could be associated with the excitation of certain degrees of freedom characteristic of quantum gravity (which we might want to describe as ``virtual black holes''). Such excitations would, in turn, generate modifications of the Schrödinger evolution equation in essentially all situations. Could this be the source of the collapse events in collapse theories? Could there be a unified picture in which loss of information in black holes is accounted for by similar features as those occurring in collapse theories? Bellow we show how these questions can be answered in the affirmative.\footnote{See \cite{P11,P13} for extended discussions of all these issues.} Before doing so, though, we need to say a few words about density matrices. 

It is important to distinguish between two different types of density matrices that are often confused, but have very different \emph{physical} interpretations. On the one hand, density matrices are used to describe either \emph{ensembles} or situations in which one only has statistical information regarding the actual pure state of a system;\footnote{We are assuming, together with the standard interpretation, that individual isolated systems always possess pure quantum states represented by individual rays in the Hilbert space (or corresponding objects in the algebraic approach).} these cases are refereed to as \emph{mixed} states or \emph{proper} mixtures. On the other hand, density matrices are used for the description of a subsystem of an isolated quantum system and are constructed by taking the trace over the rest of the system of the pure density matrix of the full system; density matrices of this type are called \emph{reduced} density matrices or \emph{improper} mixtures (see, e.g., \cite{Espagnat}). It is therefore important to distinguish between proper and improper thermal states: proper thermal states describe either ensembles with states distributed thermally or a single system for which we lack information regarding its true pure state, but know that it is distributed thermally; improper thermal states, in contrast, describe a single subsystem, which is entangled with the rest of the system in such a way that its reduced density matrix is thermal (e.g., the Minkowski vacuum, described in the Rindler wedge, using Rindler modes adapted to the boost Killing field, after tracing over the rest of spacetime).

Going back to black holes, what Hawking's calculation shows is that the initially pure state of the quantum field evolves into a final one which, when tracing over the interior region of the black hole, becomes a reduced or improper thermal state. The important question, then, is how to interpret such a final state when the black hole evaporates completely, so there is no longer an interior region to trace over? If, with us, one assumes that information gets lost during the process, then what one needs to do is to explain how the initial pure state transforms into a proper or mixed thermal state when the black hole evaporates. 

Below we explain how this transformation from a pure state into a proper thermal state, during the black hole evaporation process, can be achieved by incorporating objective collapses into the standard picture. As in \cite{P10}, we do so in a toy model based on the following simplifications:
\begin{enumerate}
\item We assume that the black hole evaporation leaves no remnant.
\item We work with a simple 2D black hole know as the Callan-Giddings-Harvey-Strominger (CGHS) model, where most calculations can be done explicitly.
\item We use a non-relativistic toy version of CSL naively adapted to quantum field theory in curved spacetime.
\item We make a few natural simplifying assumptions about what happens when quantum gravity cures the black hole singularity.
\end{enumerate}
Moreover, as we explain below in detail (see also section \ref{GRandC}), we work under the assumption that the CSL collapse parameter is not fixed, but increases with the local curvature. In the next section we a give sketch of our proposal (see \cite{P10} for more details).
%%%%%%%%%%%%%%%%%%%%%%%%%%%%%%% 
\subsubsection{Black holes and objective collapses}
%%%%%%%%%%%%%%%%%%%%%%%%%%%%%%% 
We begin with a brief review of the CGHS black hole (\cite{CGHS}), which is a very convenient 2D toy model in order to study black hole formation and evaporation. The starting point is the action
\begin{equation}
{ S=\frac{1}{2\pi}\int d^2x\sqrt{-g}\left[e^{-2\phi}\left[R+4(\nabla \phi)^2+4\Lambda^2\right]-\frac{1}{2}(\nabla f)^2\right],}
\end{equation}
where $R$ is the Ricci scalar for the metric $g_{ab}$, $\phi$ is a dilaton field, $\Lambda$ is a cosmological constant and $f$ is a scalar field characterizing matter (note that the dilaton is included in the gravity sector). The CGHS solution for the metric (see Figure 1) is given by
\begin{equation}
ds^2 = -\frac{dx^{+}dx^{-}}{-\Lambda^2x^{+}x^{-}-(M/\Lambda x_{0}^{+})(x^{+}-x_{0}^{+})\Theta(x^{+}-x_{0}^{+})} ,
%\\
\end{equation}
and corresponds to a null shell of matter collapsing gravitationally along the world line $x^+ = x^+_0$, leading to the formation of a black hole. That is, in regions $I$ and $I'$ the metric is Minkowskian, but in regions $II$ and $III$ it represents a black hole (with region $II$ its exterior and region $III$ its interior).
\begin{figure}[h]
\centering
\includegraphics[natwidth=.,natheight=.8, scale=.15]{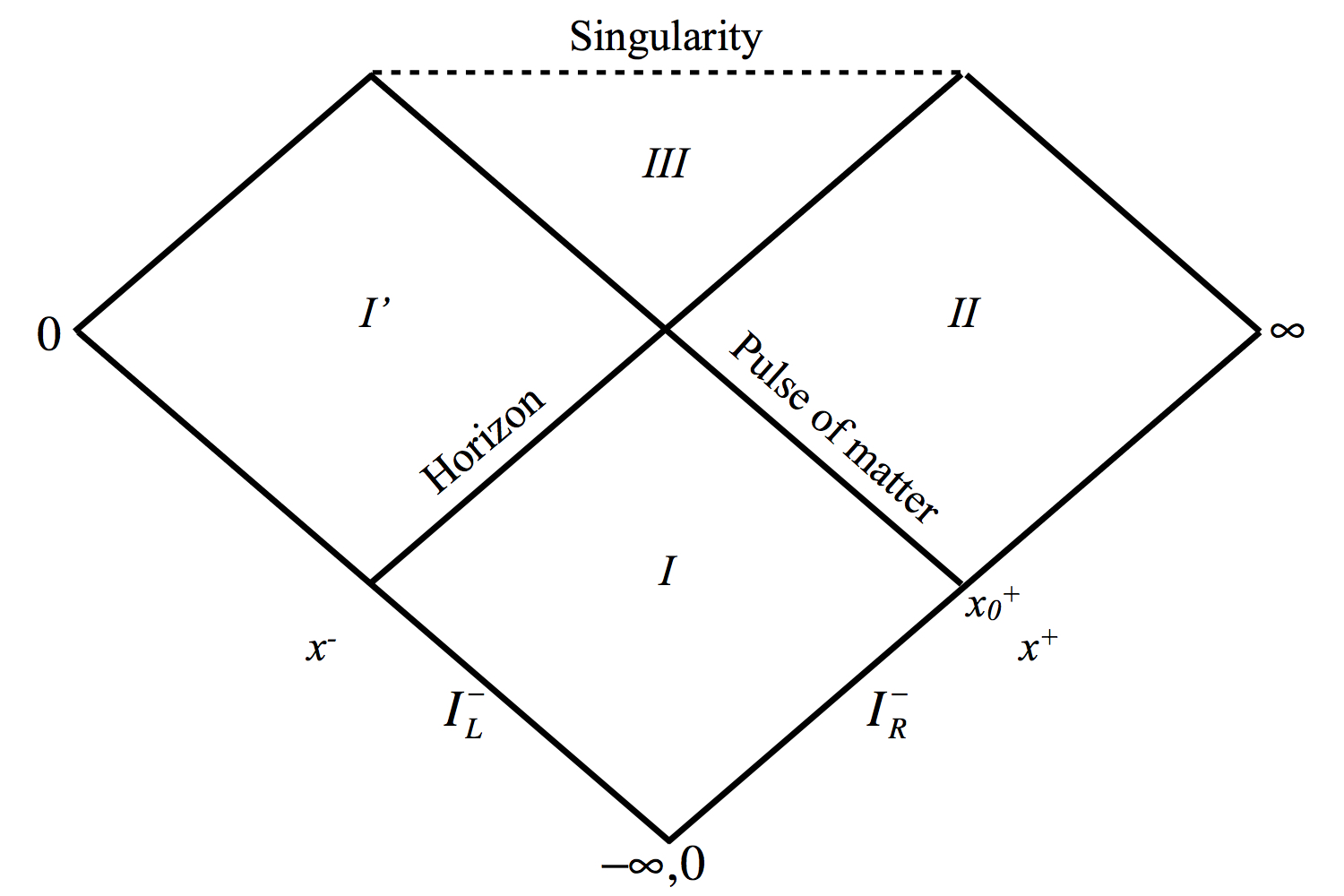} 
 \caption{Penrose diagram for the CGHS model.}
\end{figure}

Next we consider quantum aspects of the model. Given our semi-classical approach and the fact that, in the CGHS model, the matter sector is represented only by $f$, only such object has to be treated quantum mechanically. We take the null past asymptotic regions $I^{-}_{L}$ and $I^{-}_{R}$ as the \emph{in} region and both the interior and exterior of the black hole as the \emph{out} region. Regarding the Bogolubov transformations from \emph{in} to \emph{out} modes, it turns out that the only non-trivial ones, and, in fact, those that account for the Hawking radiation, are the transformations from \emph{in} to \emph{exterior} modes. In particular, for the initial state, which is taken to be the vacuum for the right moving modes and the left moving pulse that forms the black hole, we have
\begin{equation}
\label{IS}
\kets{\Psi_{in}} = \kets{0_{in}}_{R}\otimes \kets {Pulse}_{L} = N \displaystyle\sum_{\alpha} C_{F_{\alpha }} \kets{F_{\alpha }}^{ext} \otimes \kets{F_{\alpha }}^{int}\otimes \kets{Pulse}_{L}
\end{equation}
where the $\kets{F_{\alpha }}$ represent states with a \emph{finite} number of particles, $N$ is a normalization constant and the $C_{F_{\alpha }}$ are determined by the Bogolubov transformations (see \cite{Fab:05}). Note that it is at this stage that one normally takes a trace over the interior degrees of freedom to find an improper thermal state that corresponds to the Hawking radiation. We will \emph{not} do so here because we want to provide a description of the complete system.

We finally consider modifications to the evolution of the system arising from a dynamical reduction theory such as CSL. The idea is to work in the interaction picture, treating the CSL modifications, formally, as an interaction. Therefore, we take the free evolution as that encoded in the quantum field operator, while the state of the field evolves as a result of the modified dynamics. In order to use CSL, we need a foliation. For this we use hypesurfaces with constant Ricci curvature $R$ in the inside, which match suitable hypesurfaces in the outside (see details in \cite{P10}). We also introduce a ``foliation time'' parameter $\tau$ to label these hypersurfaces.
 
The next step is to select a suitable CSL collapse operator, for which we judiciously choose one that drives initial states into states of definite number of particles in the inside region (note that the CSL equations can be generalized to drive collapse into a state of a joint eigen-basis of a set of commuting operators). We also incorporate the assumption that the CSL collapse mechanism is amplified by the curvature of spacetime. We do so by postulating the rate of collapse $\lambda$ to depend on the local curvature as in equation (\ref{Wd}); however, in this toy model, instead of Weyl scalar we use the Ricci scalar. We thus take
\begin{equation}
{\lambda (R) = \lambda_0 \left[1+\left(\frac{R}{\mu}\right)^\gamma\right]}
\end{equation}
where $\lambda_0$ is the flat space collapse rate, $R$ is the Ricci scalar of the CGHS spacetime, $\gamma >1$ is a constant and $\mu$ provides an appropriate scale. As a result, in the region of interest (i.e., the black hole interior, where the CSL modifications can become large), we effectively have $\lambda=\lambda(\tau)$. The point of this modification is to ensure for the resulting evolution to drive the initial state into an eigenstate of the collapse operator in a \emph{finite} time (and not an infinite one as in standard CSL).

With all these assumptions, we can characterize the effect of CSL on the initial state in equation (\ref{IS}) as simply driving it to one of the eigenstates of the joint number operator. Thus, at an hypersurfaces of constant $\tau$ located very close to the singularity, the state of the field is
 \begin{equation}
\kets{\Psi_{in, \tau}} = %N C_{F_{\tilde{\alpha}}}
 \kets{F_{\tilde{\alpha}}}^{ext} \otimes \kets{F_{\tilde{\alpha} }}^{int}\otimes \kets{Pulse}_{L} ,
\end{equation}
where $\tilde{\alpha}$ labels the particular state of definite number of particles randomly chosen by the CSL dynamics to collapse the initial state into.

The next ingredient we need to incorporate in our treatment is the role of quantum gravity. That we do by making a couple of mild assumptions regarding the nature of the underlying quantum gravity theory. First, we assume that it resolves the singularity of the black hole and that it leads, after the complete evaporation, to some reasonable region of spacetime that can be described, to a good approximation, by classical notions. Second, we assume that quantum gravity does not lead, at the macro-scale, to large violations of the basic spacetime conservation laws. With these assumptions, we can further evolve the interior state to a \emph{post-singularity} ($ps$) stage an write
\begin{equation}
\kets{\Psi_{in, ps}} = %N \displaystyle C_{F_{\tilde{\alpha}}}
 \kets{F_{\tilde{\alpha}}}^{ext} \otimes \kets{0^{ps}} , 
\end{equation}
where $\kets{0^{ps}}$ represents a zero energy and momentum state, corresponding to a trivial region of spacetime (recall that we have ignored possible small remnants). We end up, then, with a pure quantum state. However, we do not know which one it is because that is determined randomly by the specific realization of the stochastic functions of the CSL dynamics.
 
What we have to do at this point, then, is to consider an ensemble of systems, all prepared in the same initial state of equation (\ref{IS}). Then, we consider the CSL evolution of the ensemble up to the hypersurface just before the singularity and, finally, we use what was assumed about quantum gravity to further evolve those states to a post-singularity stage. As a result of all this, together with the fact that the CSL evolution at the ensemble level leads to probability distributions that are compatible with Born's rule, one can check that the density matrix characterizing the ensemble in the post-singularity stage is
\begin{eqnarray}
\rho^{ps} &=& N^2 \sum_{\alpha} e^{-\frac{2\pi}{\Lambda} E_\alpha} \kets{F_\alpha}^{ext} \otimes \kets{0^{ps}} \bras{F_\alpha}^{ext} \otimes \bras{0^{ps}} \nonumber \\
 &=& \rho^{ext}_{Thermal} \otimes \kets{0^{ps}} \bras{0^{ps}} .
\end{eqnarray}
Thus, the starting point is a pure state and at the end the situation is described by a proper thermal state, which expresses the fact that, as a result of the stochastic character of the CSL evolution, we only have statistical information regarding the actual final pure state of the system. Information was lost as a result of the (slightly modified) quantum evolution. Clearly, there is nothing paradoxical in the resulting picture. 

One might harbor various concerns about the overall picture we have presented. The first one is the dependence of the results on the choice of foliation and, more generally, the fact that the whole scheme employed is non-relativistic. Of course this is a serious problem, but it can be overcome by employing a relativistic version of the collapse dynamics. In fact, a similar analysis was performed in \cite{P12} using a relativistic version of CSL recently developed in \cite{Bed}. We also note that our choice of collapse operators was \emph{ad hoc} and one might worry whether the end result crucially depends on that choice. It is important to point out, thought, that the no-signaling theorem (also valid within GRW and CSL) ensures that the density matrix characterizing the situation in the region exterior to the horizon is insensitive to the choice of collapse operators relevant for the dynamics in the interior.

There are other aspects in which the above treatment needs improvement. One is the question of back-reaction, which we have not, at this point, incorporated in any meaningful sense (beyond the simple expectation that as the black hole radiates its mass decreases). The other mayor concern is the question of the nature of the universal collapse operators that should appear in the general form of the collapse dynamics. Such theory has to involve a universal collapse operator that in the non-relativistic context reduces to a smeared position operator and in the situation at hand corresponds to something leading to similar results as those obtained in the treatment above. Finally, we have made some important assumptions about quantum gravity that, of course, could simply turn out to be false.
%%%%%%%%%%%%%%%%%%%%%%%%%%%%%%%%%%%%%%%%%%%%%%%%%%%%%%%%%%%%%% 
\subsection{The problem of time in quantum gravity}
\label{Time}
%%%%%%%%%%%%%%%%%%%%%%%%%%%%%%%%%%%%%%%%%%%%%%%%%%%%%%%%%%%%%% 
The diffeomorphism invariance of general relativity leads to very problematic implications when one attempts to apply a canonical quantization procedure to the theory. In order to implement such method, one starts with the Hamiltonian formulation of the classical theory, which corresponds to choosing a foliation $\Sigma_t$ and taking as canonical data the 3-metrics $h_{ab}$ of the hypersurfaces and their conjugate momenta $\pi^{ab}$ (for simplicity we will restrict ourselves to general relativity in the absence of matter fields). The foliation is then characterized in terms of the so-called lapse function $N$ and shift vectors $N^a$, which determine the points of the $\Sigma_t$ and $\Sigma_{ t +\Delta t}$ hypersurfaces used to describe the evolution of the canonical variables. Moreover, the canonical data is constrained by
\begin{equation} 
 {\cal H} (h_{ab}, \pi^{ab}) =0 \quad \text{and} \quad {\cal H}_a (h_{ab}, \pi^{ab}) =0
 \end{equation}
where ${\cal H}$ and ${\cal H}_a $ are specific functions of the canonical variables. These are known as the Hamiltonian and diffeomorphism constraints, respectively. Finally, the Hamiltonian that generates the evolution along the vector field $t^a= n^a N + N^a$ can be expressed as 
 \begin{equation} \label{Hamiltonian}
 H =\int d^3 x \sqrt h [ N {\cal H} + N^a {\cal H}_a ].
 \end{equation}
 
The canonical quantization procedure involves replacing the canonical variables $h_{ab}$ and $\pi^{ab}$ by operators in a Hilbert space $\hat h_{ab}$ and $ \hat \pi^{ab}$, such that the Poisson brackets are suitably replaced by commutation relations. In order to impose the constraints, one starts with an auxiliary Hilbert space ${\mathcal H}_{Aux}$ (usually taken to be the space of wave functionals on the configuration variables), from which the physical Hilbert space ${\mathcal H}_{Phys}$ is constructed as the subset of ${\mathcal H}_{Aux}$ satisfying the operational constrains
 \begin{equation} \label{quantum constrains}
 \hat {\cal H }\Psi ( h_{ab}) =0 \quad \text{and} \quad
 \hat {\cal H}_a \Psi ( h_{ab}) =0.
 \end{equation}

As always, time evolution is controlled by the Schr\"{o}dinger equation, but since, on physical states, $ \hat H =\int d^3 x \hat {\sqrt{h}} [ N \hat {\cal H} + N^a \hat {\cal H}_a ]=0$, it is clear that the state of the system is independent of $t$. We end up, then, with physical states that may depend on the spatial metric $h_{ab}$ but not on time. Time has then completely disappeared for the physical description provided by the theory; this is the problem of time in quantum gravity.

The problem of time, then, is related to the fact that the Hamiltonian, which is the generator of change, is equal to zero. The situation clearly changes when collapse processes are incorporated into the picture, that is, if the standard evolution equation is replaced by something like
\begin{equation} \label{Schroedinger2}
 i d\Psi ( h_{ab}) = \left\lbrace \int dt \int d^3 x \hat {\sqrt{h}} [ N \hat {\cal H} + N^a \hat {\cal H}_a ] + \int d^4 x \hat {\cal C}(x) \right\rbrace \Psi ( h_{ab}) ,
 %=\int d^3 x \sqrt h [ N {\cal H} + N^a {\cal H}_a ] ,
 \end{equation}
where $ \hat {\cal C}(x) $ is an operator characterizing the effects of the collapse dynamics. In such case, the full evolution will not be controlled exclusively by a Hamiltonian constructed out of the theory's constraints so, even though the constraints annihilate the physical states, they will display a non-trivial evolution. The upshot is that temporal change is brought back into the picture, thus solving the problem of time.

%%%%%%%%%%%%%%%%%%%%%%%%%%%%%%%%%%%%%%%%%%%%%%%%%%%%%%%%%%%%%% 
\subsection{The Weyl curvature hypothesis}
 \label{PH}
%%%%%%%%%%%%%%%%%%%%%%%%%%%%%%%%%%%%%%%%%%%%%%%%%%%%%%%%%%%%%% 
The second law of thermodynamics has generated intense debates throughout the years. The main issue under discussion is the fact that it is not clear how such a time-asymmetric law can emerge from fundamental laws of nature, like those of general relativity and quantum theory, which are essentially time-symmetric. A popular solution to such question, the so-called \emph{past hypothesis}, is to postulate that the universe started in a state of extremely low entropy. In \cite{Pen:79}, Penrose conjectured that the past hypothesis arises from a constraint on the initial value of the Weyl curvature, keeping it very low. Penrose's proposal is very different from standard physical laws, which govern the dynamics, rather than the initial conditions, and some have found this odd feature dissatisfying (e.g., lacking explanatory power). Here we show how the adoption of an objective collapse model, particularly one with a curvature-dependent collapse rate as in equation (\ref{Wd}), allows for a \emph{dynamical} explanation of Penrose's conjecture (see \cite{P14} for more details). 

We begin by assuming that the very early universe was characterized by wildly varying, generically high, values of the Weyl curvature $W$. As a result, the collapse rate $\lambda (W)$ was very large and the evolution was dominated by the stochastic component of CSL (i.e., the non-standard term in equation (\ref{CSL1}), with $\lambda_0$ substituted by $\lambda (W)$). That implied an extremely stochastic evolution for both matter and geometry. Such type of evolution continued until, by mere chance, a small value of $W$ was obtained. From then on, the evolution settled into the Hamiltonian dominated regime (i.e., the standard part of equation (\ref{CSL1})), associated with an almost constant value of the matter density, a value of $R$ with similar characteristics and a very small value of $W$. Of course, such scenario is precisely what Penrose conjectured. Therefore, a collapse model with a curvature-dependent collapse rate offers a dynamical justification for what Penrose introduced as a constraint on initial conditions in order to account for the very special initial state of the universe.

After a small value of $W$ is randomly achieved, the appropriate conditions for the onset of inflation develop. Inflation further flattens the spatial geometry and leads to the standard story we described in section \ref{CI}, with collapses playing a crucial role in the production of seeds of structure. After that, the rate of collapse diminishes, but not to zero, and becomes essential for the job it was originally designed for, i.e., the suppression of superpositions of well-localized ordinary macroscopic objects. The regime where $W$ is large is encountered again within black holes where, as we explained in section \ref{BH}, takes care of the information loss required in order to avoid paradoxical situations.
%%%%%%%%%%%%%%%%%%%%%%%%%%%%%%%%%%%%%%%%%%%%%%%%%%%%%%%%%%%%%%
\section{Conclusions}
\label{D}
%%%%%%%%%%%%%%%%%%%%%%%%%%%%%%%%%%%%%%%%%%%%%%%%%%%%%%%%%%%%%% 
We have reviewed some known arguments, and offered some additional ones, suggesting a deep connection between the conceptual problems at the foundations of quantum theory and topics that are usually considered to belong to cosmology and quantum gravity research. More particularly, we have shown that one of the proposals for dealing with the measurement problem in quantum mechanics, namely the dynamical reduction approach, offers attractive resolutions of various outstanding problems that occur at the interface of the quantum and general relativistic realms. These include the origin of the seeds of cosmic structure from quantum fluctuations during the inflationary era, the black hole information puzzle and the problem of time in quantum theory. 

Moreover, we have seen that, following such line of thought, one also finds natural explanations for the lack of detection of primordial gravity waves (trough the search for B-modes of polarization in the CMB), for the peculiar value of the cosmological constant and for some of the special features that, according to Penrose, must have characterized the initial state of the universe. Of course, all these results involve, at this point, several simplifying considerations and some {\it ad hoc} assumptions, so, in order to be considered as completely satisfactory answers, they require a detailed elucidation of some of the underlying premises. However, we find that the progress achieved so far must be regarded as rather promising, in particular when considering the short time the program has been under development and the scarce number of researchers involved. We are convinced that, by having a well-defined program in which to work, problems and obstacles become visible; and by facing them, one moves forward either by ruling out specific alternatives or by pinpointing crucial characteristics that can make them viable.
%%%%%%%%%%%%%%%%%%%%%%%%%%%%%%%%%%%%%%%%%%%%%%%%%%%%%%%%%%%%%% 
%%%%%%%%%%%%%%%%%%%%%%%%%%%%%%%%%%%%%%%%%%%%%%%%%%%%%%%%%%%%%%
\section*{Acknowledgments}
%%%%%%%%%%%%%%%%%%%%%%%%%%%%%%%%%%%%%%%%%%%%%%%%%%%%%%%%%%%%%%
We acknowledge partial financial support from DGAPA-UNAM project IG100316. DS was further supported by CONACyT project 101712.
%%%%%%%%%%%%%%%%%%%%%%%%%%%%%%%%%%%%%%%%%%%%%%%%%%%%%%%%%%%%%%
%%%%%%%%%%%%%%%%%%%%%%%%%%%%%%%%%%%%%%%%%%%%%%%%%%%%%%%%%%%%%%
%%%%%%%%%%%%%%%%%%%%%%%%%%%%%%%%%%%%%%%%%%%%%%%%%%%%%%%%%%%%%%
\bibliographystyle{plainnat}
\bibliography{biblioCUP}

\begin{thebibliography}{91}
\providecommand{\natexlab}[1]{#1}
\providecommand{\url}[1]{\texttt{#1}}
\expandafter\ifx\csname urlstyle\endcsname\relax
  \providecommand{\doi}[1]{doi: #1}\else
  \providecommand{\doi}{doi: \begingroup \urlstyle{rm}\Url}\fi

\bibitem[Albert(1996)]{Alb:96}
D.~Z. Albert.
\newblock Elementary quantum metaphysics.
\newblock In J.~T. Cushing, A.~Fine, and S.~Goldstein, editors, \emph{Bohmian
  Mechanics and Quantum Theory: An Appraisal}, pages 277--284. Kluwer Academic
  Publishers, 1996.

\bibitem[Allori(2015)]{Allori}
V.~Allori.
\newblock Primitive ontology in a nutshell.
\newblock \emph{Int. J. Quant. Found.}, 1\penalty0 (3):\penalty0 107--122,
  2015.

\bibitem[Almheiri et~al.(2013)Almheiri, Marolf, Polchinski, and
  Sully]{firewalls}
A.~Almheiri, D.~Marolf, J.~Polchinski, and J.~Sully.
\newblock Black holes: complementarity or firewalls?
\newblock \emph{JHEP}, 62, 2013.

\bibitem[Bassi et~al.(2013)Bassi, Lochan, Satin, Singh, and Ulbricht]{Bassi}
A.~Bassi, K.~Lochan, S.~Satin, T.~Singh, and H.~Ulbricht.
\newblock Models of wave-function collapse, underlying theories, and
  experimental tests.
\newblock \emph{Rev. Mod. Phys.}, 85:\penalty0 471, 2013.

\bibitem[Bedingham et~al.(2014)Bedingham, Dürr, Ghirardi, Goldstein, Tumulka,
  and Zanghi]{RevMat}
D.~Bedingham, D.~Dürr, G.~C. Ghirardi, S.~Goldstein, R.~Tumulka, and
  N.~Zanghi.
\newblock Matter density and relativistic models of wave function collapse.
\newblock \emph{J. Stat. Phys.}, 154:\penalty0 623–631, 2014.

\bibitem[Bedingham et~al.(2016)Bedingham, Modak, and Sudarsky]{P12}
D.~Bedingham, S.~K. Modak, and D.~Sudarsky.
\newblock Relativistic collapse dynamics and black hole information loss.
\newblock \emph{Phys. Rev. D}, 94\penalty0 (4):\penalty0 045009, 2016.

\bibitem[Bedingham(2011)]{Bed}
D.~J. Bedingham.
\newblock Relativistic state reduction dynamics.
\newblock \emph{Found. Phys.}, 41:\penalty0 686, 2011.

\bibitem[Bell(1981)]{Bel:81}
J.~S. Bell.
\newblock Quantum mechanics for cosmologists.
\newblock In \emph{Quantum Gravity II}. Oxford University Press, 1981.

\bibitem[Bell(1987)]{Bel:87}
J.~S. Bell.
\newblock Are there quantum jumps?
\newblock In C.~W. Kilminster, editor, \emph{Schrödinger: Centenary
  Celebration of a Polymath}, page 109–123. Cambridge University Press, 1987.

\bibitem[Bell(2004)]{BellB}
J.~S. Bell.
\newblock \emph{Speakable and Unspeakable in Quantum Mechanics}.
\newblock Cambridge University Press, 2nd edition, 2004.

\bibitem[Bohm(1952)]{Bohm}
D.~Bohm.
\newblock A suggested interpretation of quantum theory in terms of `hidden'
  variables.
\newblock \emph{Phys. Rev.}, 85:\penalty0 166--193, 1952.

\bibitem[Bohr(1949)]{Bohr}
N.~Bohr.
\newblock Discussion with einstein on epistemological problems in atomic
  physics.
\newblock In P.~A. Schilpp, editor, \emph{Albert Einstein:
  Philosopher-scientist}, page 201–241. La Salle: Open Court, 1949.

\bibitem[Bonder et~al.(2015)Bonder, Okon, and Sudarsky]{Bon:15}
Y.~Bonder, E.~Okon, and D.~Sudarsky.
\newblock Can gravity account for the emergence of classicality?
\newblock \emph{Phys. Rev. D}, 92:\penalty0 124050, 2015.

\bibitem[Bonder et~al.(2016)Bonder, Okon, and Sudarsky]{Bon:16}
Y.~Bonder, E.~Okon, and D.~Sudarsky.
\newblock Questioning universal decoherence due to gravitational time dilation.
\newblock \emph{Nature Phys.}, 12:\penalty0 2, 2016.

\bibitem[Callan et~al.(1992)Callan, Giddings, Harvey, and Strominger]{CGHS}
C.~G. Callan, S.~B. Giddings, J.~A. Harvey, and A.~Strominger.
\newblock Evanescent black holes.
\newblock \emph{Phys. Rev. D}, 45:\penalty0 R1005, 1992.

\bibitem[Carlip(2008)]{Carlip}
S.~Carlip.
\newblock Is quantum gravity necessary?
\newblock \emph{Class. Quant. Grav.}, 25:\penalty0 154010, 2008.

\bibitem[Cañate et~al.(2013)Cañate, Pearle, and Sudarsky]{P7}
P.~Cañate, P.~Pearle, and D.~Sudarsky.
\newblock Csl quantum origin of the primordial fluctuation.
\newblock \emph{Phys. Rev. D}, 87:\penalty0 104024, 2013.

\bibitem[Colella et~al.(1975)Colella, Overhauser, and Werner]{COW}
R.~Colella, A.~W. Overhauser, and S.~A. Werner.
\newblock Observation of gravitationally induced quantum interference.
\newblock \emph{Phys. Rev. Lett.}, 34:\penalty0 1472–1474, 1975.

\bibitem[Derakhshani(2014)]{Der:14}
M.~Derakhshani.
\newblock Newtonian semiclassical gravity in the ghirardi–rimini–weber
  theory with matter density ontology.
\newblock \emph{Phys. Lett. A}, 378:\penalty0 14--15, 2014.

\bibitem[d'Espagnat(1976)]{Espagnat}
B.~d'Espagnat.
\newblock \emph{Conceptual Foundations of Quantum Mechanics}.
\newblock Addison-Wesley, 2nd edition, 1976.

\bibitem[Diez-Tejedor and Sudarsky(2012)]{P6}
A.~Diez-Tejedor and D.~Sudarsky.
\newblock Towards a formal description of the collapse approach to the
  inflationary origin of the seeds of cosmic structure.
\newblock \emph{JCAP}, 045:\penalty0 1207, 2012.

\bibitem[Diosi(1984)]{Diosi1}
L.~Diosi.
\newblock Gravitation and quantum-mechanical localization of macro-objects.
\newblock \emph{Phys. Lett. A}, 105:\penalty0 199, 1984.

\bibitem[Diosi(1987)]{Diosi2}
L.~Diosi.
\newblock A universal master equation for the gravitational violation of
  quantum mechanics.
\newblock \emph{Phys. Lett. A}, 120:\penalty0 377, 1987.

\bibitem[Ellis et~al.(1989)Ellis, Mohanty, and Nanopoulos]{Ell:89}
J.~Ellis, S.~Mohanty, and D.~V. Nanopoulos.
\newblock Quantum gravity and the collapse of the wavefunction.
\newblock \emph{Phys. Lett. B}, 221:\penalty0 113--119, 1989.

\bibitem[Eppley and Hannah(1977)]{EH}
K.~Eppley and E.~Hannah.
\newblock The necessity of quantizing the gravitational field.
\newblock \emph{Found. Phys.}, 7:\penalty0 51–68, 1977.

\bibitem[Everett(1957)]{Eve:57}
H.~Everett.
\newblock `relative state' formulation of quantum mechanics.
\newblock \emph{Rev. Mod. Phys.}, 29\penalty0 (3), 1957.

\bibitem[Fabbri and Navarro-Salas(2005)]{Fab:05}
A.~Fabbri and J.~Navarro-Salas, editors.
\newblock \emph{Modeling Black Hole Evaporation}.
\newblock Imperial College Press, 2005.

\bibitem[Gambini et~al.(2004)Gambini, Porto, and Pullin]{GP}
R.~Gambini, R.~A. Porto, and J.~Pullin.
\newblock A relational solution to the problem of time in quantum mechanics and
  quantum gravity: a fundamental mechanism for quantum decoherence.
\newblock \emph{New J. Phys.}, 5:\penalty0 45, 2004.

\bibitem[Gell-Mann and Hartle(1990)]{CH3}
M.~Gell-Mann and J.~B. Hartle.
\newblock Quantum mechanics in the light of quantum cosmology.
\newblock In W.~Zurek, editor, \emph{Complexity, Entropy, and the Physics of
  Information, SFI Studies in the Sciences of Complexity, Vol. VIII}. Addison
  Wesley, 1990.

\bibitem[Ghirardi et~al.(1986)Ghirardi, Rimini, and Weber]{GRW}
G.~C. Ghirardi, A.~Rimini, and T.~Weber.
\newblock Unified dynamics for microscopic and macroscopic systems.
\newblock \emph{Phys. Rev. D}, 34:\penalty0 470--491, 1986.

\bibitem[Ghirardi et~al.(1995)Ghirardi, Grassi, and Benatti]{Ghi:95}
G.~C. Ghirardi, R.~Grassi, and F.~Benatti.
\newblock Describing the macroscopic world: closing the circle within the
  dynamical reduction program.
\newblock \emph{Found. Phys.}, 35:\penalty0 5, 1995.

\bibitem[Goldstein and Teufel(2001)]{Gol.Teu:01}
S.~Goldstein and S.~Teufel.
\newblock Quantum spacetime without observers: ontological clarity and the
  conceptual foundations of quantum gravity.
\newblock In C.~Callender and N.~Huggett, editors, \emph{Physics Meets
  Philosophy at the Planck Scale}, pages 275--289. Cambidge University Press,
  2001.

\bibitem[Griffiths(1984)]{CH1}
R.~B. Griffiths.
\newblock Consistent histories and the interpretation of quantum mechanics.
\newblock \emph{J. Stat. Phys.}, 36:\penalty0 219, 1984.

\bibitem[Hartle(2006)]{Hartle-Cosmology}
J.~B. Hartle.
\newblock Generalizing quantum mechanics for quantum gravity.
\newblock \emph{Int. J. Theor. Phys.}, 45:\penalty0 1390--1396, 2006.

\bibitem[Hawking(1975)]{Haw:75}
S.~Hawking.
\newblock Particle creation by black holes.
\newblock \emph{Commun. Math. Phys.}, 43:\penalty0 199–220, 1975.

\bibitem[Hu and Verdaguer(2008)]{Hu-Verdaguer}
B.~L. Hu and E.~Verdaguer.
\newblock Stochastic gravity: Theory and applications.
\newblock \emph{Liv. Rev. Rel.}, 11:\penalty0 3, 2008.

\bibitem[Huggett and Callender(2001)]{CH}
N.~Huggett and C.~Callender.
\newblock Why quantize gravity (or any other field for that matter)?
\newblock \emph{Phil. Sci.}, 68\penalty0 (3):\penalty0 S382--S394, 2001.

\bibitem[Jacobson(1995)]{JacobsonEmergent}
T.~Jacobson.
\newblock Thermodynamics of spacetime: The einstein equation of state.
\newblock \emph{Phys. Rev. Lett.}, 75:\penalty0 1260, 1995.

\bibitem[Josset et~al.(2016)Josset, Perez, and Sudarsky]{Josset:2016}
T.~Josset, A.~Perez, and D.~Sudarsky.
\newblock Dark energy as the weight of violating energy conservation.
\newblock \emph{Phys. Rev. Lett.}, ?:\penalty0 ?, 2016.

\bibitem[Kamionkowski and Kovetz(2016)]{Kamionkowski}
M.~Kamionkowski and E.~D. Kovetz.
\newblock The quest for b modes from inflationary gravitational waves.
\newblock \emph{Annual Rev. Astron. and Astroph.}, 54:\penalty0 227--269, 2016.

\bibitem[Knuth and Bahreyni(2014)]{SorkinEmergent}
K.~H. Knuth and N.~Bahreyni.
\newblock A potential foundation for emergent space-time.
\newblock \emph{J. Math. Phys.}, 55:\penalty0 112501, 2014.

\bibitem[Kuo and Ford(1993)]{Kuo}
C.~I. Kuo and L.~Ford.
\newblock Semiclassical gravity theory and quantum fluctuations.
\newblock \emph{Phys. Rev. D}, 47:\penalty0 4510, 1993.

\bibitem[Landau et~al.(2012)Landau, Scoccola, and Sudarsky]{P5}
S.~J. Landau, C.~G. Scoccola, and D.~Sudarsky.
\newblock Cosmological constraints on nonstandard inflationary quantum collapse
  models.
\newblock \emph{Phys. Rev. D}, 85:\penalty0 123001, 2012.

\bibitem[León and Sudarsky(2010)]{P2}
G.~León and D.~Sudarsky.
\newblock The slow roll condition and the amplitude of the primordial spectrum
  of cosmic fluctuations: Contrasts and similarities of standard account and
  the `collapse scheme'.
\newblock \emph{Class. Quant. Grav.}, 27:\penalty0 225017, 2010.

\bibitem[León et~al.(2011)León, De~Unanue, and Sudarsky]{P4}
G.~León, A.~De~Unanue, and D.~Sudarsky.
\newblock Multiple quantum collapse of the inflaton field and its implications
  on the birth of cosmic structure.
\newblock \emph{Class. Quant. Grav.}, 28:\penalty0 155010, 2011.

\bibitem[León et~al.(2013)León, Landau, Scoccola, and Sudarsky]{P8}
G.~León, S.~J. Landau, C.~G. Scoccola, and D.~Sudarsky.
\newblock Quantum origin of the primordial fluctuation spectrum and its
  statistics.
\newblock \emph{Phys. Rev. D}, 88:\penalty0 023526, 2013.

\bibitem[Liddle(1999)]{Liddle}
A.~R. Liddle.
\newblock An introduction to cosmological inflation.
\newblock In A.~Masiero, G.~Senjanovic, and A.~Smirnov, editors, \emph{High
  Energy Physics and Cosmology}, 1999.

\bibitem[Majhi et~al.(2016)Majhi, Okon, and Sudarsky]{P16}
A.~Majhi, E.~Okon, and D.~Sudarsky.
\newblock Reassessing the link between b-modes and inflation.
\newblock \emph{arXiv:1607.03523}, 2016.

\bibitem[Maldacena and Susskind(2013)]{EREPR}
J.~Maldacena and L.~Susskind.
\newblock Cool horizons for entangled black holes.
\newblock \emph{Fortsch. Phys.}, 61:\penalty0 781--811, 2013.

\bibitem[Mattingly(2005)]{Mat1}
J.~Mattingly.
\newblock Is quantum gravity necessary?
\newblock In A.~J. Kox and J.~Eisenstaedt, editors, \emph{The Universe of
  General Relativity}, pages 325--338. Birkhäuser, 2005.

\bibitem[Mattingly(2006)]{Mat2}
J.~Mattingly.
\newblock Why epply and hannah's thought experiment fails.
\newblock \emph{Phys. Rev. D}, 73:\penalty0 064025, 2006.

\bibitem[Mielnik(1974)]{Mie:74}
B.~Mielnik.
\newblock Generalized quantum mechanics.
\newblock \emph{Comm. Math. Phys.}, 35:\penalty0 221--256, 1974.

\bibitem[Modak et~al.(2015)Modak, Ortiz, Peña, and Sudarsky]{P10}
S.~K. Modak, L.~Ortiz, I.~Peña, and D.~Sudarsky.
\newblock Non-paradoxical loss of information in black hole evaporation in a
  quantum collapse model.
\newblock \emph{Phys. Rev. D}, 91\penalty0 (12):\penalty0 124009, 2015.

\bibitem[Mortonson and Seljak(2014)]{nullB-modes1}
M.~J. Mortonson and U.~Seljak.
\newblock A joint analysis of planck and bicep2 b modes including dust
  polarization uncertainty.
\newblock \emph{JCAP}, 1410:\penalty0 035, 2014.

\bibitem[Nesvizhevsky et~al.(2002)Nesvizhevsky, Börner, Petukhov, Abele,
  Baeßler, Rueß, Stöferle, Westphal, Gagarsky, Petrov, and
  Strelkov]{neutrons}
V.~V. Nesvizhevsky, H.~G. Börner, A.~K. Petukhov, H.~Abele, S.~Baeßler, F.~J.
  Rueß, Th. Stöferle, A.~Westphal, A.~M. Gagarsky, G.~A. Petrov, and A.~V.
  Strelkov.
\newblock Quantum states of neutrons in the earth's gravitational field.
\newblock \emph{Nature}, 415:\penalty0 297, 2002.

\bibitem[Ney and Albert(2013)]{WF}
A.~Ney and D.~Z. Albert, editors.
\newblock \emph{The Wave Function: Essays on the Metaphysics of Quantum
  Mechanics}.
\newblock Oxford University Press, 2013.

\bibitem[Nimmrichter and Hornberger(2015)]{Nim.Hor:15}
S.~Nimmrichter and K.~Hornberger.
\newblock Stochastic extensions of the regularized schrödinger-newton
  equation.
\newblock \emph{Phys. Rev. D}, 91:\penalty0 024016, 2015.

\bibitem[Okon and Sudarsky(2014{\natexlab{a}})]{CHus1}
E.~Okon and D.~Sudarsky.
\newblock On the consistency of the consistent histories approach to quantum
  mechanics.
\newblock \emph{Found. Phys.}, 44:\penalty0 19–33, 2014{\natexlab{a}}.

\bibitem[Okon and Sudarsky(2014{\natexlab{b}})]{CHus2}
E.~Okon and D.~Sudarsky.
\newblock Measurements according to consistent histories.
\newblock \emph{Stud. Hist. Phil. Mod. Phys.}, 48:\penalty0 7–12,
  2014{\natexlab{b}}.

\bibitem[Okon and Sudarsky(2014{\natexlab{c}})]{P9}
E.~Okon and D.~Sudarsky.
\newblock Benefits of objective collapse models for cosmology and quantum
  gravity.
\newblock \emph{Found. Phys.}, 44:\penalty0 114--143, 2014{\natexlab{c}}.

\bibitem[Okon and Sudarsky(2015{\natexlab{a}})]{CHus3}
E.~Okon and D.~Sudarsky.
\newblock The consistent histories formalism and the measurement problem.
\newblock \emph{Stud. Hist. Phil. Mod. Phys.}, 52:\penalty0 217–222,
  2015{\natexlab{a}}.

\bibitem[Okon and Sudarsky(2015{\natexlab{b}})]{P11}
E.~Okon and D.~Sudarsky.
\newblock The black hole information paradox and the collapse of the wave
  function.
\newblock \emph{Found. Phys.}, 44:\penalty0 461--470, 2015{\natexlab{b}}.

\bibitem[Okon and Sudarsky(2016{\natexlab{a}})]{P13}
E.~Okon and D.~Sudarsky.
\newblock Black holes, information loss and the measurement problem.
\newblock \emph{Found. Phys.}, 2016{\natexlab{a}}.

\bibitem[Okon and Sudarsky(2016{\natexlab{b}})]{P14}
E.~Okon and D.~Sudarsky.
\newblock A (not so?) novel explanation for the very special initial state of
  the universe.
\newblock \emph{Class. Quant. Grav.}, 2016{\natexlab{b}}.

\bibitem[Okon and Sudarsky(2016{\natexlab{c}})]{P15}
E.~Okon and D.~Sudarsky.
\newblock Less decoherence and more coherence in quantum gravity, inflationary
  cosmology and elsewhere.
\newblock \emph{Found. Phys.}, 46:\penalty0 852–879, 2016{\natexlab{c}}.

\bibitem[Omnès(1994)]{CH2}
R.~Omnès.
\newblock \emph{Interpretation of Quantum Mechanics}.
\newblock Princeton University Press, 1994.

\bibitem[Page and Geilker(1981)]{Page}
D.~N. Page and C.~D. Geilker.
\newblock Indirect evidence for quantum gravity.
\newblock \emph{Phys. Rev. Lett.}, 47:\penalty0 979, 1981.

\bibitem[Peacock(1999)]{Pea}
J.~A. Peacock.
\newblock \emph{Cosmological Physics}.
\newblock Cambridge University Press, 1999.

\bibitem[Pearle(1989)]{CSL}
P.~Pearle.
\newblock Combining stochastic dynamical state vector reduction with
  spontaneous localization.
\newblock \emph{Phys. Rev. A}, 39:\penalty0 2277--2289, 1989.

\bibitem[Pearle(2015)]{PearleRel}
P.~Pearle.
\newblock Relativistic dynamical collapse model.
\newblock \emph{Phys. Rev. D}, 91\penalty0 (10):\penalty0 105012, 2015.

\bibitem[Pearle and Squires(1994)]{Mass-dep2}
P.~Pearle and E.~Squires.
\newblock Bound state excitation, nucleon decay experiments, and models of wave
  function collapse.
\newblock \emph{Phys. Rev. Lett.}, 73:\penalty0 1, 1994.

\bibitem[Pearle and Squires(1996)]{Mass-dep1}
P.~Pearle and E.~Squires.
\newblock Gravity, energy conservation, and parameter values in collapse
  models.
\newblock \emph{Found. Phys.}, 26:\penalty0 291--305, 1996.

\bibitem[Penrose(1979)]{Pen:79}
R.~Penrose.
\newblock Singularities and time-asymmetry.
\newblock In S.~W. Hawking and W.~Israel, editors, \emph{General Relativity: An
  Einstein Centenary Survey}, page 581–638. Cambridge University Press, 1979.

\bibitem[Penrose(1981)]{Penrose1}
R.~Penrose.
\newblock Time asymmetry and quantum gravity.
\newblock In C.~J. Isham, R.~Penrose, and D.~W. Sciama, editors, \emph{Quantum
  Gravity II}. Clarendon Press, 1981.

\bibitem[Penrose(1996)]{Penrose2}
R.~Penrose.
\newblock On gravity's role in quantum state reduction.
\newblock \emph{Gen. Rel. Grav.}, 28:\penalty0 581, 1996.

\bibitem[Percival(1995)]{Per:95}
I.~C. Percival.
\newblock Quantum space-time fluctuations and primary state diffusion.
\newblock \emph{Proc. Roy. Soc. London Ser. A}, 451:\penalty0 503, 1995.

\bibitem[Perez et~al.(2006)Perez, Sahlmman, and Sudarsky]{P1}
A.~Perez, H.~Sahlmman, and D.~Sudarsky.
\newblock On the quantum mechanical origin of the seeds of cosmic structure.
\newblock \emph{Class. Quant. Grav.}, 23:\penalty0 2317, 2006.

\bibitem[Perlmutter et~al.(1999)]{Perlmutter:1998np}
S.~Perlmutter et~al.
\newblock Measurements of omega and lambda from 42 high redshift supernovae.
\newblock \emph{Astrophys. J.}, 517:\penalty0 565--586, 1999.

\bibitem[Pikovski et~al.(2015)Pikovski, Zych, Costa, and Brukner]{Pik:15}
I.~Pikovski, M.~Zych, F.~Costa, and C.~Brukner.
\newblock Universal decoherence due to gravitational time dilation.
\newblock \emph{Nature Phys.}, 11:\penalty0 668, 2015.

\bibitem[{Planck Collaboration}(2015)]{Adam:2015rua}
{Planck Collaboration}.
\newblock Planck 2015 results. i. overview of products and scientific results.
\newblock \emph{Astron. Astrophys}, 594:\penalty0 A1, 2015.

\bibitem[{Planck Collaboration}(2016)]{nullB-modes2}
{Planck Collaboration}.
\newblock E- and b-modes of dust polarization from the magnetized filamentary
  structure of the interstellar medium.
\newblock \emph{Astron. Astrophys.}, 586:\penalty0 A141, 2016.

\bibitem[Riess et~al.(1998)]{Riess:1998cb}
A.~G. Riess et~al.
\newblock Observational evidence from supernovae for an accelerating universe
  and a cosmological constant.
\newblock \emph{Astron. J.}, 116:\penalty0 1009--1038, 1998.

\bibitem[Ryu and Takayanagi(2006)]{Ryu:06}
S.~Ryu and T.~Takayanagi.
\newblock Holographic derivation of entanglement entropy from the anti–de
  sitter space/conformal field theory correspondence.
\newblock \emph{Phys. Rev. Lett.}, 96:\penalty0 181602, 2006.

\bibitem[Saunders et~al.(2010)Saunders, Barrett, Kent, and Wallace]{MW}
S.~Saunders, J.~Barrett, A.~Kent, and D.~Wallace, editors.
\newblock \emph{Many Worlds? Everett, Quantum Theory, and Reality}.
\newblock Oxford University Press, 2010.

\bibitem[Seiberg(2006)]{SeibergEmergent}
N.~Seiberg.
\newblock Emergent spacetime.
\newblock \emph{arXiv:hep-th/0601234}, 2006.

\bibitem[Sorkin(1993)]{Sorkin}
R.~D. Sorkin.
\newblock Impossible measurements on quantum fields.
\newblock In B.~L. Hu and T.~A. Jacobson, editors, \emph{Directions in General
  Relativity, Vol. II: a Collection of Essays in honor of Dieter Brill's
  Sixtieth Birthday}. Cambridge University Press, 1993.

\bibitem[Sudarsky(2011)]{P3}
D.~Sudarsky.
\newblock Shortcomings in the understanding of why cosmological perturbations
  look classical.
\newblock \emph{IJMPD}, 20:\penalty0 509, 2011.

\bibitem[Sudarsky(2014)]{Prague}
D.~Sudarsky.
\newblock The inflationary origin of the seeds of cosmic structure: Quantum
  theory and the need for novel physics.
\newblock \emph{Fund. Theor. Phys.}, 177:\penalty0 349, 2014.

\bibitem[Tilloy and Diosi(2016)]{DiosiNew}
A.~Tilloy and L.~Diosi.
\newblock Sourcing semiclassical gravity from spontaneously localized quantum
  matter.
\newblock \emph{Phys. Lett. D}, 93:\penalty0 024026, 2016.

\bibitem[Tumulka(2006)]{Tumulka}
R.~Tumulka.
\newblock A relativistic version of the ghirardi-rimini-weber model.
\newblock \emph{J. Stat. Phys.}, 125\penalty0 (821):\penalty0 10, 2006.

\bibitem[Weinberg(1989)]{Weinberg:1988cp}
S.~Weinberg.
\newblock The cosmological constant problem.
\newblock \emph{Rev. Mod. Phys.}, 61:\penalty0 1--23, 1989.

\end{thebibliography}
%%%%%%%%%%%%%%%%%%%%%%%%%%%%%%%%%%%%%%%%%%%%%%%%%%%%%%%%%%%%%%
\end{document}